\documentclass[%
aip,
amsmath,amssymb,
reprint,%
]{revtex4-2}
\usepackage{graphicx}
\usepackage{dcolumn}
\usepackage{bm}
\usepackage{hyperref}
\hypersetup{colorlinks=true,linkcolor=blue,citecolor=blue,filecolor=blue,urlcolor=blue}
\usepackage{color}
\usepackage{xcolor,soul}
\usepackage{alltt}
\usepackage{braket}
\usepackage{algcompatible}
\usepackage{newfloat}
\usepackage{tikz}
\usepackage{subcaption}
\usepackage{float}
\usepackage{comment}
\usepackage{soul}
\usepackage[font=footnotesize, labelfont=bf]{caption}

\usepackage{xargs}

\begin{document}
\author{Mohammed Alhissi}
\email{mohammed-k-m.alhissi@uni-konstanz.de}
\affiliation{Fachbereich Physik, Universit\"at Konstanz, 78464 Konstanz, Germany}

\author{Andreas Zumbusch}
\email{andreas.zumbusch@uni-konstanz.de}
\affiliation{Fachbereich Chemie, Universit\"at Konstanz, 78464 Konstanz, Germany}

\author{Matthias Fuchs}
\email{matthias.fuchs@uni-konstanz.de}
\affiliation{Fachbereich Physik, Universit\"at Konstanz, 78464 Konstanz, Germany}

\date{\today}

\title{Observation of Liquid Glass in Molecular Dynamics Simulations}

\begin{abstract}
Molecular anisotropy plays an important role in the glass transition of a liquid. Recently, a novel bulk glass state has been discovered by optical microscopy experiments on suspensions of ellipsoidal colloids. 'Liquid glass' is a disordered analog of a nematic liquid crystal, in which rotation motion is hindered but particles diffuse freely. Global nematic order is suppressed as clusters of aligned particles intertwine. We perform Brownian dynamics simulations to test the structure and dynamics of a dense system of soft ellipsoidal particles. As seen in experiments and in accordance with predictions from mode coupling theory, on the time scale of our simulations rotation motion is frozen but translation motion persists in liquid glass. Analyses of the dynamic structure functions for translation and rotation  corroborates the presence of two separate glass transitions for rotation and translation, respectively. Even though the equilibrium state should be a nematic, aligned structures remain small and orientational order rapidly decays with increasing size. Long-wavelength  fluctuations  are remnants of the isotropic-nematic transition. 
\end{abstract}

\maketitle

\section{Introduction}
\label{s:Intro}

Colloids are ubiquitous in our everyday lives. Additionally, they play an important role as models, where they can be considered as 'big atoms' \cite{Poon2004}. To date, most work exploring the structure and dynamics of colloidal suspensions has been performed with spherical particles. However, many physical and technological colloidal systems are non-spherical. Over the last two decades, the role of colloid form has come into the focus of scientific interest \cite{Dugyala2013,Barrat2024}. Shape matters especially at higher colloid concentrations, where the packing gets more intricate because of steric constraints. This is apparent already for ellipsoidal particles that are the simplest generalization of spherical bodies. For elliposidal colloids, theory and simulations predict a rich phase diagram of equilibrium fluid, crystal \cite{Pfleiderer2007} and nematic phases \cite{Odriozola2012}, plus additionally jammed  \cite{Donev2004,Hoy2022,Rocks2023} and  glass states \cite{Letz2000, DeMichele2007, Pfleiderer2008, Chong2005} due to the presence of structural and orientational correlations. The 3D structure of fluid dispersions of ellipsoids has been studied with confocal microscopy\cite{Cohen}.

Because the colloidal glass transition features many of the phenomena found in atomic and molecular systems, colloids have been used extensively as model systems to study the glass transition  \cite{Weeks2017,Gokhale2016,Stillinger2013}. For a two dimensional system, glassy behavior in the orientation dynamics of anisotropic fluid was detected using event driven molecular dynamics simulations \cite{xu15,wang18b}. In three dimensions, decoupling between the translation and rotation degrees of freedoms was observed in molecular dynamics simulations of a binary fluid composed of dumbbell-shaped particles where the rotation relaxation times were found to be smaller than the translation ones \cite{Chong}. Using confocal microscopy of suspensions of prolate ellipsoidal colloids with an aspect ratio $\eta=3.5$, we could recently show the importance of colloidal shape on the thermal, quiescent glass \cite{RollerLaganapan2020}. Liquid glass as a new glass state, predicted by theory \cite{Letz2000} twenty years ago, was discovered. It is formed as intermediate material in an intriguing two-step vitrification process when increasing the volume fraction. Measurements of angular and translation time-dependent correlation functions demonstrated that when the packing density increases, the orientational motion freezes before the translation one. In between these two glass transitions, a liquid glass exists as a state where colloids can diffuse but do not de-correlate their local angular alignment. It was found that the global nematic order vanishes because ramified clusters of aligned particles interpenetrate and hinder each other \cite{RollerLaganapan2020}.

Glass states of ellipsoidal particles have been described in mode coupling theory (MCT) by Schilling and coworkers \cite{Schilling1997,Schilling2000,Letz2000,Schilling2002}. MCT derives the structural arrest from the equilibrium static structure factors as input, and for larger aspect ratios predicted two different glass transitions for rotation and translation motion. The topology of state space was further elucidated by developing a schematic model of MCT \cite{RollerLaganapan2020}.  It captures the generic coupling of two sets of degrees of freedom, translation and rotation ones, and captures their frequency-dependent interrelation in retarded friction kernels. The model predicts two generic MCT glass transitions, and a region in between that corresponds to the existence range of liquid glass. MCT makes specific predictions on the relaxation of the incipient glassy structure in the supercooled state \cite{Goetze}, that we will use to identify liquid glass in the present simulation work. 

The formation of liquid glass is driven by the directional correlations accompanying nematic ordering. Yet, in some systems, these directional correlations preempt the ordering transition \cite{RollerLaganapan2020}, while in other systems, glass formation causes texture dynamics to arrest in nematic states. Highly elongated colloidal rods, for instance, show arrested space-dependent director fields in suspensions far in the nematic regime \cite{Kang2013a,Kang2013b}. Static glassy structures obtained by driving particles in external fields \cite{Crassous2014,Pal2018,Pal2020a,Pal2022b}, and jammed states in athermal systems of ellipsoids \cite{Hoy2022,Rocks2023} are other instructive examples for the importance of local alignment in amorphous solids made from anisotropic particles. A liquid glass-like state also exists in two-dimensional colloidal films \cite{Zheng2014,Zheng2011, Mishra2014,Sokolowsky2016,Wang2022}, where it forms via the mutual steric hindrance of compact nematic domains. The nematic order of the two-dimensional domains extends over hundreds of particles, while it decays to zero fast in the three-dimensional liquid glass  \cite{RollerLaganapan2020}. This suggests that ramified and fractal nematic precursors are the building blocks of liquid glass in bulk dispersions. Its mechanical properties should be intriguing as it is expected to flow like a liquid but to transmit torques like a solid even though it lacks nematic order. 

Here, we report results of Brownian dynamics simulations shedding light on the microscopic structural correlations and the transport properties of liquid glass. The study is motivated by the intriguing absence of a clear observation of liquid glass formation in earlier simulations. To model the hard interaction potential of the ellipsoidal colloids found experimentally, event driven Brownian dynamics (EDBD) simulations \cite{Donev2004,Donev2005,DeMichele2006,Scala2007,DeMichele2010} were performed previously \cite{DeMichele2007,RollerLaganapan2020}. In these cases, only rather small systems of around 500 ellipsoidal particles could be simulated because of the numerical complexity to sample random motion without tolerating overlaps of anisotropic particles. In the two-dimensional simulations, an artificial dynamics using kinetic Monte Carlo (kMC) was employed for similar reasons \cite{Zheng2011}. The EDBD simulations reported the formation of a nematic phase as expected from equilibrium Monte Carlo studies. The formation of a nematic state could be suppressed by rough walls, which led to the formation of a (regular) glass \cite{RollerLaganapan2020}, viz.~the state of arrested diffusion and rotation. The isotropic-nematic transition prevented the observation of a clear two-step relaxation in angular correlation functions as expected by MCT for liquid glass formation, and the identification of the two different glass transitions for rotation and translation required appreciable extrapolation of the rotation and translation diffusion coefficients to zero \cite{DeMichele2007}. Similar extrapolations of the relaxation times of angular and translation functions was necessary in the kMC simulations in order to indicate the two glass transitions in the films \cite{Zheng2011}. No EDBD or kMC simulations have been reported in the liquid glass state, however. This has prevented simulations to test whether the dynamics in this state exhibits the decoupling of rotation and translation and follows the predictions of MCT. To overcome the limitations of the previous simulations, we first employ techniques that allow to simulate far larger ellipsoid systems than previously studied.  Second, to be able to quench beyond the nematic state,we develop a soft repulsive potential that enables large steps in the effective volume fraction. It simplifies the established Gay-Berne potential \cite{GayBerne1981} by only keeping the repulsion as has been done for spherical particles. For a test how close this approaches true hard interactions see e.g.~Ref. \cite{Lange2009}. A second aspect neglected in the previous EDBD simulations of ellipsoidal colloids \cite{Donev2004,Donev2005,DeMichele2006,Scala2007,DeMichele2010,DeMichele2007,RollerLaganapan2020} concerns the anisotropic diffusion of the particles parallel and perpendicular to their main axis.  In our work, we use the Brownian dynamics algorithm developed in Ref.~\onlinecite{Ilie-briels} and take the rotation and translation diffusion coefficients from measurements of the liquid-glass forming colloids in dilute solution \cite{RollerLaganapan2020}.

The paper is organized as follows. In Sect.~\ref{s:Models} the system as specified by its interaction potential is introduced. Sect.~\ref{s:Definition} summarizes the definitions of the structural and dynamic functions that are evaluated.  In Sect.~\ref{s:Sim}, aspects of the simulation methodology are given, and the central section, Sect.~\ref{s:Results} gives the simulation results and compares them with experimental data where available. First, the isotropic and the nematic phases are characterized, and in Sect.~\ref{ss:liquid glass} the liquid glass is discussed.  Sect.~\ref{s:MCT} compares with MCT predictions and analyses the intermediate time dynamics of translation and angular correlation functions. The conclusions and a discussion of the results are given in Sect.~\ref{s:Conclusion}. 

\section{Model}
\label{s:Models}
We study an isotropic dispersion whose constituents are soft ellipsoidal particles with a fixed aspect ratio. The interactions of the  ellipsoids among themselves are governed by an anisotropic inverse potential. The study can build on approaches to molecular fluids, where efficient descriptions have been developed. In this section, the interaction potential is presented.

The interactions of anisotropic particles have been investigated for several decades \cite{BernePechukas1972,GayBerne1981,BerardiFavaZannoni1995,BerardiFavaZannoni1998,BerardiMuccioliZannoni2008}. In 1981, Gay and Berne modified the overlap potential \cite{BernePechukas1972} introduced by Berne and Pechukas and presented their potential which has since been known as the ``Gay-Berne potential'' \cite{GayBerne1981}. It is a generalization of the Lennard-Jones potential \cite{LennradJones1931} in that it includes orientation dependent parameters. With it, the isotropic, nematic, and the other liquid-crystalline phases could successfully be predicted in molecular dynamics simulations \cite{AdamsLuckhurstPhippen1987,ChalamGubbinsMiguelRull1991}. However, the ``Gay-Berne potential'' is limited to anisotropic uniaxial particles of the same type. Berardi, Fava, and Zannoni extended it to describe the pair interaction of two identical biaxial molecules \cite{BerardiFavaZannoni1995} and dissimilar biaxial molecules \cite{BerardiFavaZannoni1998}. Their formulation describes the orientational degrees of freedom using the Euler angles. Berardi, Muccioli and Zannoni introduced another expression of the biaxial ``Gay-Berne potential'' using quaternions to describe the orientation dynamics of the anisotropic molecules\cite{BerardiMuccioliZannoni2008}. This is more convenient for molecular dynamics simulations because the new potential with the quaternion formulation removes the singularity in the equations of motion resulting from use of the Euler angles to describe molecular rotations.

In this work we are mostly interested in an isotropic fluid at high densities. At such densities, the main forces contributing to structure formation are those due to the repulsive interactions of the particles' hard cores. In such a high-density fluid and since the system is far from the critical point, attractive interactions play negligible roles in stabilizing the fluid structure. Subsequently, we will therefore adopt a purely repulsive potential where the anisotropy enters through terms depending on the distance and the orientation of the ellipsoid. We are going to use only the \textit{repulsive} part of the Gay-Berne potential developed in Ref.\citenum{BerardiMuccioliZannoni2008}. The new potential will be referred to as a \textit{``Repulsive Gay-Berne potential (RGB)"}. Furthermore, we are going to study a one-component anisotropic fluid made up of identical uniaxial ellipsoids with an aspect ratio $\eta=a/b$.

The RGB potential describes the pairwise repulsive interaction of two ellipsoids, $i$ and $j$, and is defined as:

\begin{equation}
\label{SRGBP}
U(\mathbf{r},\mathbf{q}^{i},\mathbf{q}^{j})= 4\epsilon_0\epsilon(\mathbf{r},\mathbf{q}^{i},\mathbf{q}^{j})\bigg[\frac{\sigma_0}{r-\sigma(\mathbf{r},\mathbf{q}^{i},\mathbf{q}^{j})+\sigma_0} \bigg]^{12}
\end{equation}
where $\epsilon_0$ and $\sigma_0$ are an energy scale and a length scale respectively. $\mathbf{r}$ is a vector connecting the centers of masses of the two ellipsoids $i$ and $j$. The 4-dimensional unit quaternion $\mathbf{q}^i=(q^i_0,q^i_1,q^i_2,q^i_3)$ specifies the orientation of the $i^{th}$ ellipsoid with the normalization condition $|\mathbf{q}^i|=1$. The anisotropic contact term $\sigma(\mathbf{r,q^i,q^j})$ and the anisotropic interaction term $\epsilon(\mathbf{r,q^i,q^j})$ are explained in detail in Ref. \citenum{BerardiMuccioliZannoni2008}. In our work, these two terms are computed by setting the side-by-side, width-to-width, and end-to-end interactions to unity for a pair of ellipsoids with aspect ratio $\eta=3.5$.

For the computation of orientation dependent quantities, transformations from the body frame into the space-fixed frame are required. The transformation defining the rotation for a unit vector from the $i^{th}$ ellipsoid body frame $\mathbf{\hat{e}^b}$ into the space fixed frame, where it becomes $\mathbf{\hat{e}^s}$, is:
\begin{equation}
 \mathbf{\hat{e}^s}=\mathbf{R^{-1}} \; \mathbf{\hat{e}^b}
\end{equation}
where $\mathbf{R}$ is the rotation matrix written in terms of the quaternion components:
\begin{equation}
\mathbf{R}=
 \begin{bmatrix}
1- 2q^2_2 - 2q^2_3 & 2q_1q_2 + 2q_0q_3 & 2q_1q_3 - 2q_0q_2\\
2q_1q_2 - 2q_0q_3  & 1-2q^2_3-2q^2_1  & 2q_2q_3 + 2q_0q_1 \\
2q_1q_3+2q_0q_2 & 2q_2q_3-2q_0q_1 & 1-2q^2_2-2q^2_1
\end{bmatrix}
\end{equation}
Note that when considering the limit of isotropic molecules, Eq. \ref{SRGBP} reduces to the repulsive part of the Lennard-Jones potential. 

\section{Definition}
\label{s:Definition}
In this section, we introduce the functions that describe the structure of our ellipsoidal fluid as well as the quantities that give us insight into the fluid's microscopic dynamics and correlations.

\subsection{Statics}
\label{ss:Statics}
For the isotropic fluid, the structure of the fluid is identified by the radial distribution function $g(\mathbf{r})$ and the isotropic structure factor $S(\mathbf{q})$ defined respectively by
\begin{equation}
 g(\mathbf{r})=\frac{1}{N\rho} \bigg\langle\sum^N_{i} \sum^N_{j}\delta(\mathbf{r} - \mathbf{r}_{ij})\bigg\rangle
\end{equation}
\begin{equation}
S(\mathbf{q})=\frac{1}{N}\big\langle \rho^*(\mathbf{q}) \rho(\mathbf{q})\big\rangle
\end{equation}
where the brackets refer to the canonical average, $\mathbf{r}_{ij}=\mathbf{r_{i} - \mathbf{r}_{j}}$ is the vector connecting the center-to-center distance between the two ellipsoids $i$ and $\beta$, and the wave vector is $\mathbf{q}=(2\pi/L) \mathbf{n}$ with $L$ being the system length size and $\mathbf{n}$ a vector of integers. $N$ is the total number of the ellipsoids in the fluid, and $\rho=(N/L^3)$ is the particle number density. $\rho(\mathbf{q})$ is the isotropic microscopic local density:
\begin{equation}
\rho(\mathbf{q})=\sum^N_{j=1} e^{i\mathbf{q.r_j}} 
\end{equation}
The functions $g(\mathbf{r})$ and $S(\mathbf{q})$ do not give any information on the orientation correlations in the spatial and wave vector domains. Therefore, in addition to these two functions, we study two quantities that describe the spatial correlations of the orientation degrees of freedom. The orientation pair distribution function $G_n(\mathbf{r})$ is defined as
\begin{equation}
 \label{G_2}
 G_n(\mathbf{r})=\frac{1}{N\rho} \bigg\langle\sum^N_{i} \sum^N_{j}\delta(\mathbf{r} - \mathbf{r}_{ij})\; P_n(\mathbf{\hat{e}_{i}} \cdot \mathbf{\hat{e}_{j}})\bigg\rangle
\end{equation}
where $\mathbf{\hat{e}_i}$ is a unit vector along the major axis of the $i^{th}$ ellipsoid expressed in a space fixed frame. $P_n$ is the Legendre polynomial of degree $n$. In the isotropic phase, $G_n(r)$ decays to zero while in the nematic and crystalline phases, it decays to nonzero values \cite{Rodriguez-Odriozola, AdamsLuckhurstPhippen1987}. More precisely, for $n=2$ this function decays to the quadratic equilibrium value of the scalar nematic order parameter  for large distances \cite{Eppenga1984} $\lim_{r\to\infty} G_2(r)=S^2\;$. 

Another interesting orientation-dependent quantity is defined in the wave vector domain. It is called the orientation structure factor \cite{Letz2000} $S_{lm}(q)$:
\begin{equation}
 \label{S_lm}
 S_{lm}({q})=\frac{1}{N} \big\langle \rho_{lm}^*(\mathbf{q}) \; \rho_{lm}(\mathbf{q}) \big\rangle
\end{equation}
$\rho_{lm}(\mathbf{q},t)$ is the orientation-dependent microscopic local density which is defined in terms of the spherical harmonics $Y_{lm}(\theta,\phi)$ discussed in Ref.~\onlinecite{Gray-Gubbins},
\begin{equation}
\rho_{lm}(\mathbf{q})=\sqrt{4\pi}\; i^l \;\sum^N_{j=1} e^{i\mathbf{q.r_j}} \; Y_{lm}(\theta,\phi)
\end{equation}
While we only consider the density auto-correlations, viz.~$S_{lm}$, off-diagonal elements have also been measured \cite{DeMichele2006}. We only compute the matrix element $S_{20}(q)$. This makes the azimuthal (the $\phi$ angle) dependent factor of $S_{20}$ disappear. The angle $\theta$ is the angle defined by the scalar product $\cos(\theta)=\mathbf{q \cdot \hat{e}_s}$ where $\mathbf{\hat{e}_s}$ is a unit vector along the long ellipsoid diameter computed in the simulation box frame. After introducing $S_{lm}({q})$ and $\rho_{lm}(\mathbf{q})$, the equilibrium static structure of our fluid can fully be investigated.

An additional important quantity is the scalar nematic order parameter $S$ which provides us with knowledge on the degree of the orientation order in our system. This scalar quantity is measured by computing the largest eigenvalue of the nematic order tensor \cite{Eppenga1984} defined as:
\begin{equation}
Q_{\alpha\beta}=\bigg \langle \frac{1}{2N} \sum^N_{j=1} \big(3 u^j_{\alpha} u^j_{\beta} - \delta_{\alpha\beta}\big) \bigg\rangle   
\end{equation}
where $u^j_{\alpha}$ is the $\alpha$ component of the eigenvector $\mathbf{\hat{u}^j}$ representing the orientation of the ellipsoid $j$ in the simulation box frame. The largest eigenvalue then is
\begin{equation}
    S=\langle P_2(\cos{\theta})\rangle
\end{equation}
where $\theta$ is the angle pointing away from the nematic director $\langle \mathbf{\hat{u}}\rangle= \frac{1}{N} \sum_j  \langle \mathbf{\hat{u}}^j\rangle$. 

\subsection{Dynamics}
\label{ss:Dynamics}
In order to gain insight into the microscopic time evolution of the system, the dynamical quantities recording translation and orientation motions in the fluid need to be defined. These dynamical functions are defined in the temporal domain and tell us inter alia whether the system is in equilibrium so that we can compute the thermodynamic quantities. The first dynamical quantity is the well-known self intermediate scattering function \cite{Hansen} $F_s(q,t)$,
\begin{equation}\label{def:fs}
 F_s(q,t)=\frac{1}{N}\bigg\langle \sum^N_{j=1} \exp\big( i\;\mathbf{q}\cdot[\mathbf{r}_j(t^{\prime}+t)-\mathbf{r}_j(t^{\prime})]\big) \bigg\rangle
\end{equation}
This function probes single particle dynamics in the fluid and gives information on \textit{self} correlation.  In $F_s$ of Eq.~\eqref{def:fs} and in all other time-dependent correlation functions, time $t'$ will be chosen large enough to ensure equilibration or aging will be discussed explicitly. No time-averaging over $t'$ will be performed. In addition to \textit{self} correlations,  \textit{cross} correlations among the ellipsoids in the fluid must be described. Therefore, we introduce the isotropic dynamic structure factor \cite{Hansen} $S(q,t)$ ,
\begin{multline}
S(\mathbf{q},t)=\frac{1}{N}\bigg\langle \sum^N_{j=1}\sum^N_{k=1} \exp\big( i\;\mathbf{q}\cdot[\mathbf{r}_j(t^{\prime}+t)-\mathbf{r}_k(t^{\prime}\big)]\big) \bigg\rangle
\end{multline}
$S(q,t)$ measures the collective dynamics and correlations of all particles in the fluid. To examine orientation-dependent dynamics and temporal correlations, the following functions are adopted. The first one is the orientation dynamic structure factor \cite{Letz2000} $S_{lm}(\mathbf{q},t)$ 
\begin{equation}
 S_{lm}(\mathbf{q},t)=\frac{1}{N} \big\langle \rho_{lm}^*(\mathbf{q},t^{\prime}) \; \rho_{lm}(\mathbf{q},t^{\prime}+t)\big\rangle
\end{equation}
and the second function is the time-dependent orientation correlation function $L_n(t)$ defined in terms of the Legendre polynomial \cite{Berne1976}
\begin{equation}
\label{L_n}
 L_n(t)=\frac{1}{N} \bigg\langle \sum^N_j P_n\big[\mathbf{\hat{e}_{j}}(t^{\prime}+t) \cdot \mathbf{\hat{e}_{j}}(t^{\prime})\big] \bigg\rangle
\end{equation}
with the same interpretation for the unit vector $\mathbf{\hat{e}_{j}}$ as in Eq.~\eqref{G_2}. The function $S_{lm}(q,t)$ measures the correlations of the coupled translation-rotation dynamics while the function $L_n(t)$ only measures the correlation of the rotation dynamics. As a special case for $n=2$, the function $L_2(t)$ decays to the quadratic value of the equilibrium  scalar nematic order parameter $S$ for large timescales \cite{Eppenga1984}, in that, $\lim_{t\to\infty} L_2(t) = S^2\;$ . Having introduced these functions, we can now probe the dynamics of the anisotropic fluid in detail.

\section{The Simulation Setup}
\label{s:Sim}
We carry out Molecular Dynamics simulations using the Large-scale Atomic/Molecular Massively Parallel Simulator "LAMMPS" \cite{LAMMPS}. The Brownian dynamics of an ellipsoidal fluid is simulated at different densities using an algorithm implemented in LAMMPS. The integrator is known in LAMMPS, as the "\textit{fix Brownian/asphere}", and is based on the work done by I. Ilie \textit{et al.} \cite{Ilie-briels} and S. Delong \textit{et al.} \cite{Delong-Balboa}. Our system of interest is composed of 2197 ellipsoids with a fixed aspect ratio $\eta=a/b=3.5$. In our Brownian simulations, the ellipsoids exist in a three dimensional box whose length size is $L$. Periodic boundary conditions are implemented in order to avoid any wall-particle interactions. The initial state of the system is prepared by uniformly distributing tiny isotropic particles on a cubic lattice. While the dynamics is integrated, the small particles are grown until they become ellipsoids with the required aspect ratio. After that, the equilibration phase was carried out taking into account the system density under study. For the isotropic states in section \ref{ss:isotropicNematicSect} the system equilibration was recognized by the decay of the translation and orientation correlation functions to zero. For the nematic states, the equilibration was run until the scalar nematic order parameter only fluctuated around its average value. Once the equilibrations for the isotropic and nematic states were achieved, the measurement windows commenced directly for each state. For the liquid glass state the system was first equilibrated into an ergodic isotropic state at the small density $\Gamma=0.18$. The liquid glass was then obtained by quenching the fluid into a high density $\Gamma > 0.31$. In order to relax the initial high forces, the system was then run until the translation correlations have relaxed. This gave the start of the measurement window for the liquid glass states in section \ref{ss:liquid glass}. In the simulations, we set $\sigma_0=2b=k_BT/\epsilon_0=1$ where $b$ is the ellipsoid short semi-axis, $k_B$ is the Boltzmann factor, and $T$ is the temperature. The dimensionless time is $t^*= t/\tau_B$ where $t$ is the time  whereas the Brownian timescale is $\tau_B=(2b)^2/\langle D_t\rangle$ with $\langle D_t\rangle$ being the average translation diffusion coefficient. We refer to the dimensionless units with the star symbol. The timesteps used in the measurement phases are $dt^*=10^{-4}$ which is used in Sect.~\ref{ss:isotropicNematicSect}, and $dt^*=5\times10^{-5}$ which is used in Sect.~\ref{ss:liquid glass}. The simulation time for  Sect.~\ref{ss:isotropicNematicSect} is $10^6 \; dt^*$ while the simulation time for Sect.~\ref{ss:liquid glass} is $60\times 10^6 \; dt^*$.  The anisotropic translation and rotation diffusion tensors input in the simulations are the body-frame diffusion tensors measured in an experiment on dilute colloids made up of hard ellipsoids. All the experimental results or data in this research are taken from the experiment done by Roller et al. \cite{RollerLaganapan2020}. It is important to note that in the short time limit, the positioning accuracy of 20 nm in the xy-direction and 50 nm in the z-direction \cite{Roller2018} and the limited statistics for long lag times contribute to the noise especially in the correlation data. The anisotropic translation diffusion tensor is $D^*_t=\text{diag}(0.937, 0.937, 1.125)$ while the anisotropic rotation diffusion tensor is $D^*_r=\text{diag}(0.123, 0.123, 0.476)$. This means that the average scalar translation diffusion is $\langle D^*_t \rangle=1$. The dimensionless distance and wavenumber are $r^*=r/\sigma_0$ and $q^*=q\sigma_0$, respectively. 

As we use soft repulsive ellipsoids in the simulation, we adopt an \textit{effective density} as parameter of the thermodynamic states. The density of the soft particles is notably temperature dependent. For the soft particles the effective density is defined as $\Gamma=\rho\sigma_0^3(\epsilon_0/k_BT)^{3/k}$. In our simulation, we set $\epsilon_0=k_BT$ and $k=12$. Using the effective density is an approximation, as the potential in Eq.~(1) has corrections to the isomorphs scaling at short distance\cite{Dyre2014}, yet $\Gamma$ captures the dominant temperature dependence. With the above setup, the simulations run at different effective densities $\Gamma$. 

Since most of the figures shown use reduced units, we are going to omit the stars from the dimensionless quantities for simplicity. When dimensional quantities are needed or discussed, their units will be stated explicitly next to them. Also, we will refer to the effective density only as the density.

\section{Results and Discussion}
\label{s:Results}
This section is divided into two parts: Identifying the equilibrium phases of the fluid (the isotropic and nematic) and finding the liquid glass as a nonequilibrium state in the soft ellipsoid fluid. For comparison with the experimental work by Roller, et al. \cite{RollerLaganapan2020},  experimental data are only available for the isotropic and the liquid glass states. 
\subsection{The Isotropic and Nematic Phases}
\label{ss:isotropicNematicSect}
\begin{figure}[h!]
     \centering
     \begin{subfigure}[b]{0.45\textwidth}
         \centering
         \includegraphics[width=\textwidth]
         {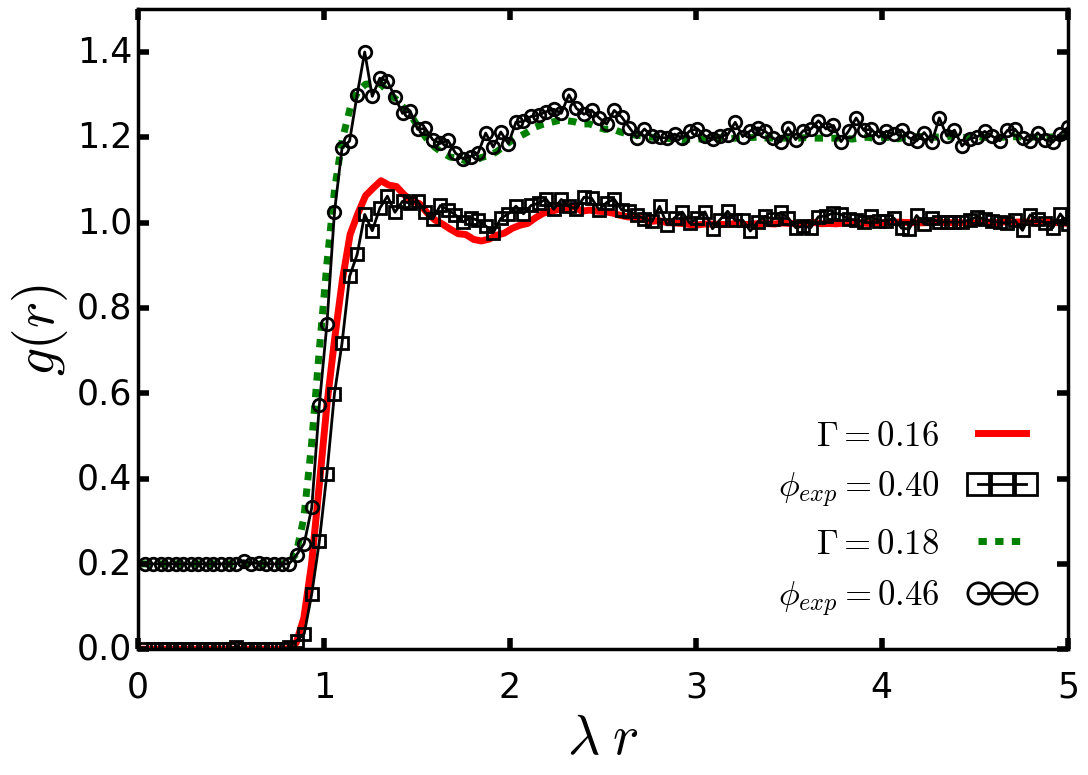}
         \caption{}
     \end{subfigure}
     \hfill
     \begin{subfigure}[b]{0.45\textwidth}
         \centering
         \includegraphics[width=\textwidth]
         {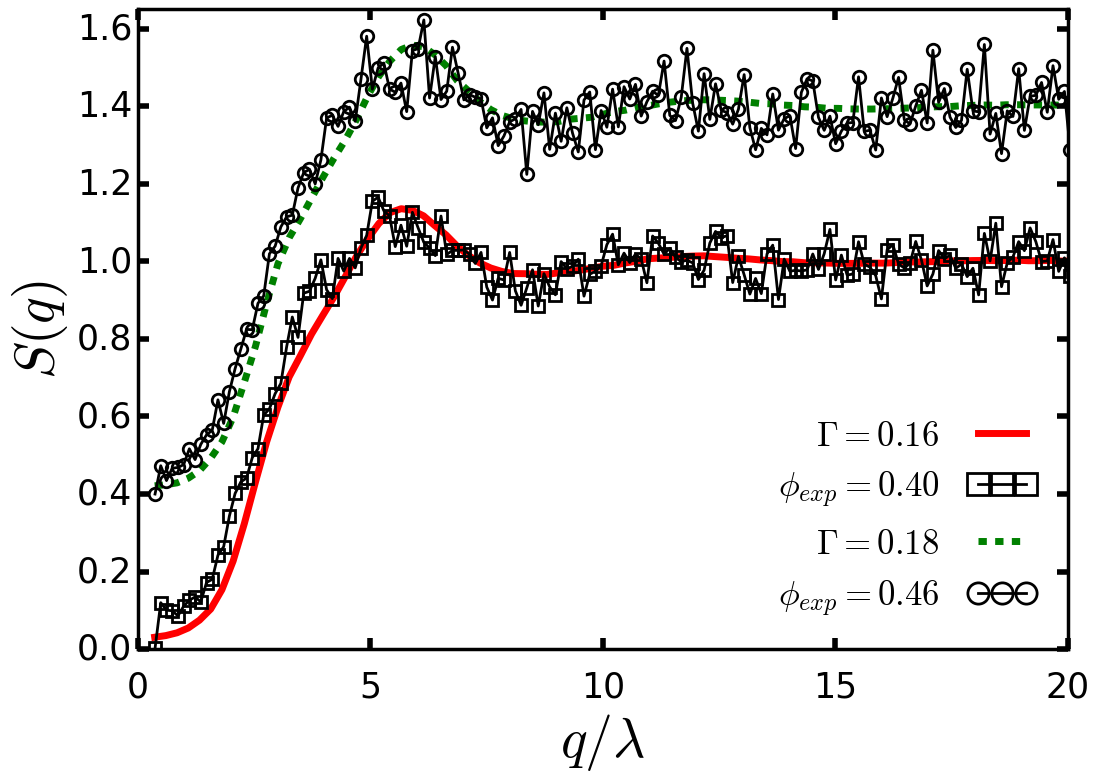}
         \caption{}
     \end{subfigure}
     \hfill
        \caption{The translation structure functions $g(r)$ and $S(q)$ for two isotropic states in the simulation (solid and dashed lines) and their isotropic counterparts in the experiment (lines marked with  squares and circles). $\phi_{exp}$ refers to the volume fractions in the experiment; one set of curves is shifted vertically for better visibility. For the experiment (hard ellipsoid) curve $\lambda=1.0$ and for the \textit{scaled} simulation (soft ellipsoid) curve $\lambda=0.83$ (see the text).}
        \label{rdfSq_isotropics}
\end{figure}
\begin{figure}[h!]
     \centering
     \begin{subfigure}[b]{0.45\textwidth}
         \centering
         \includegraphics[width=\textwidth]
         {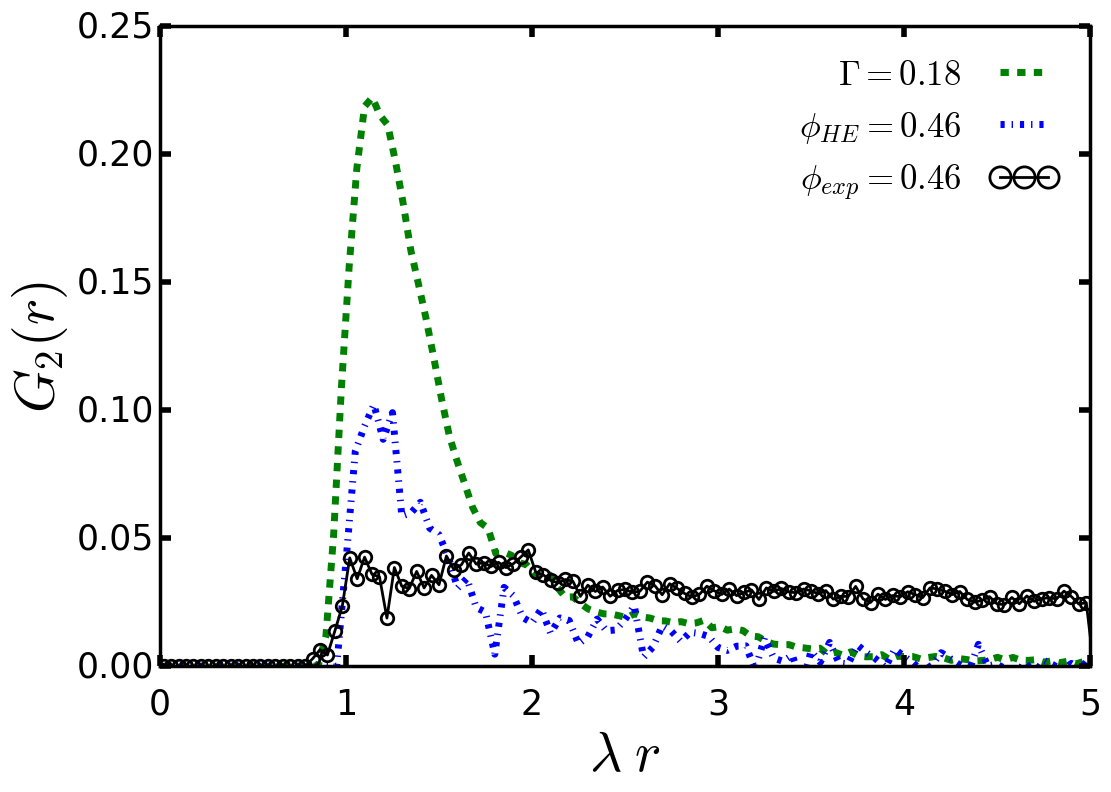}
         \caption{}
     \end{subfigure}
     \hfill
     \hfill
     \begin{subfigure}[b]{0.45\textwidth}
         \centering
         \includegraphics[width=\textwidth]
         {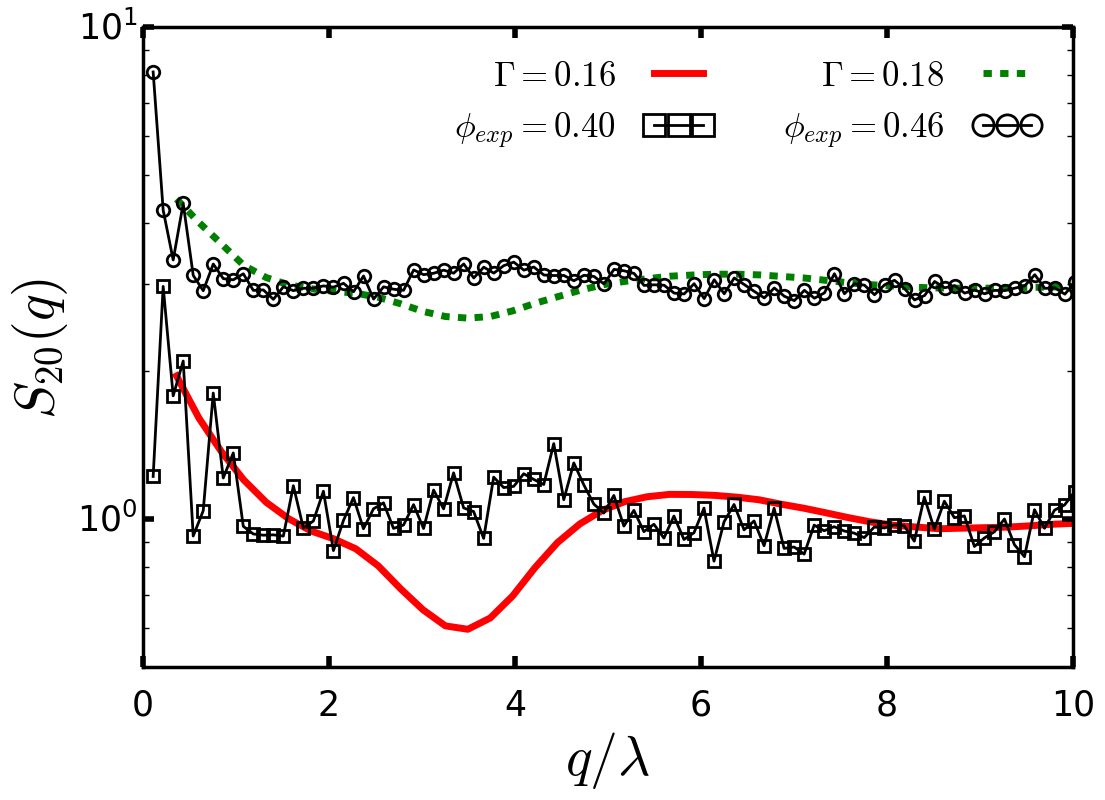}
         \caption{}
     \end{subfigure}
     \hfill
        \caption{The orientation structure functions
        $G_2(r)$ and $S_{20}(q)$ for isotropic states in the simulation (the solid and dashed lines) and their isotropic counterparts in the experiment (the lines marked with squares and circles).
        The $G_2(r)$ is shown for the isotropic state $\Gamma=0.18$ and
        $\phi_{exp}=0.46$, respectively, and results from EDBD simulations of hard ellipsoids \cite{RollerLaganapan2020} are included as blue dash-dotted line.
        The $S_{20}(q)$ is shown for two  volume fractions (one of them shifted vertically); see legends. The  $\lambda$ are the same as in Fig.~\ref{rdfSq_isotropics}.}
        \label{ordfoSq_isotropics}
\end{figure}
\begin{figure}[h!]
 \centering
 \includegraphics[width=0.45\textwidth]{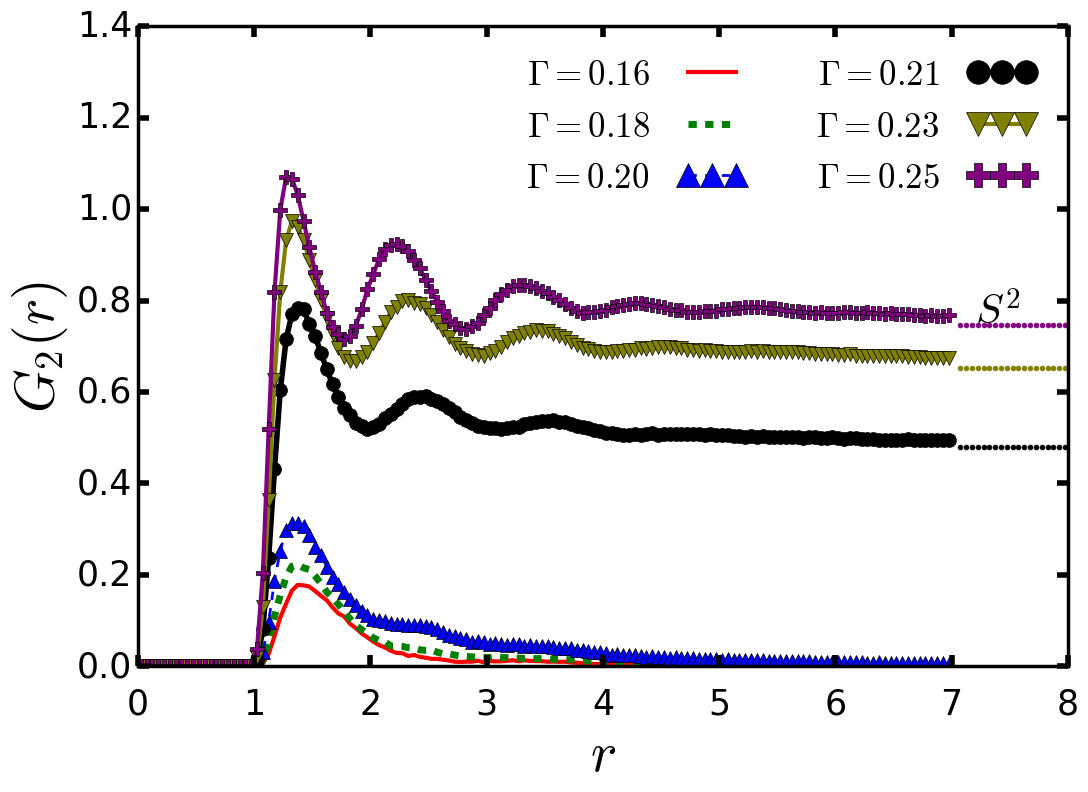}
 \caption{The simulation results for the spatial orientation correlation function $G_2(r)$ for densities given in the legend. It decays to zero and does not show a long range orientation order in the isotropic phase for $\Gamma < 0.21$. In the nematic phase, this function shows stronger orientation order that can be seen in the peaks at multiples of $2b$. Also it does not decay to zero, but rather approaches the limit of $S^2$ for large distances; {here $S$ is taken from Fig.~\ref{clusterAnalysis} (at particle number $\approx 275$), see text for details}.}
 \label{ordfIsoNem}
\end{figure}
\begin{figure}[h!]
     \centering
     \begin{subfigure}[b]{0.45\textwidth}
         \centering
         \includegraphics[width=\textwidth]
         {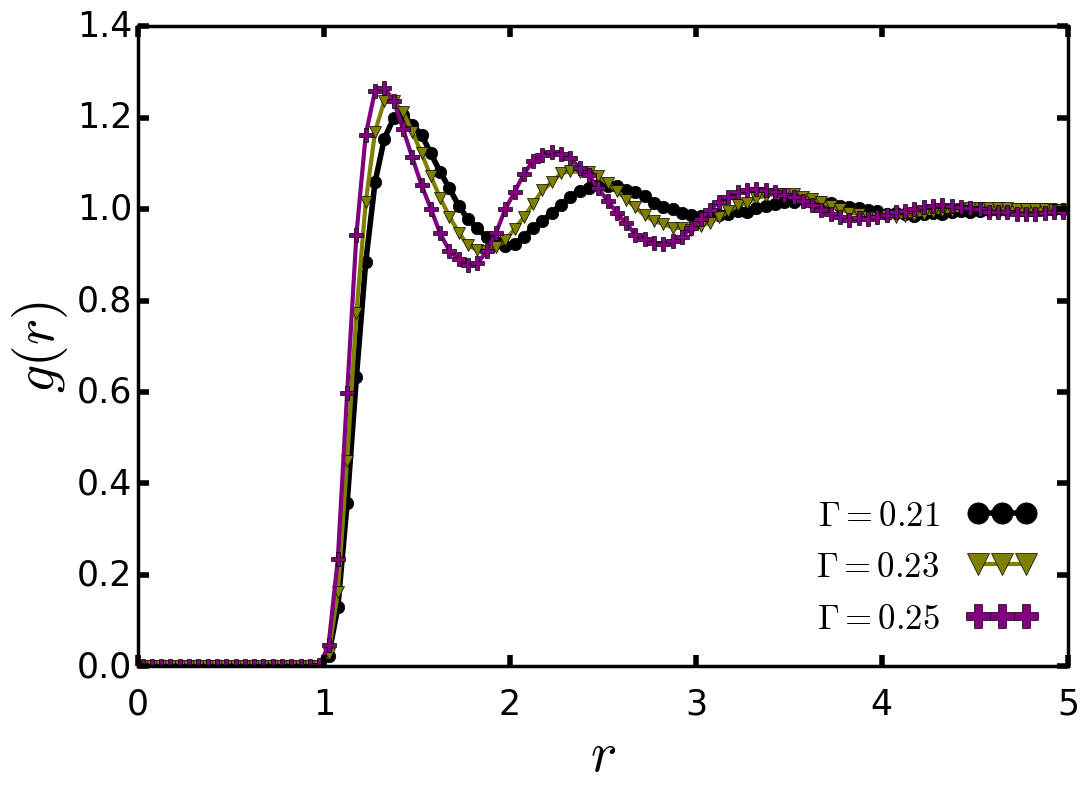}
         \caption{}
     \end{subfigure}
     \hfill
     \begin{subfigure}[b]{0.45\textwidth}
         \centering
         \includegraphics[width=\textwidth]
         {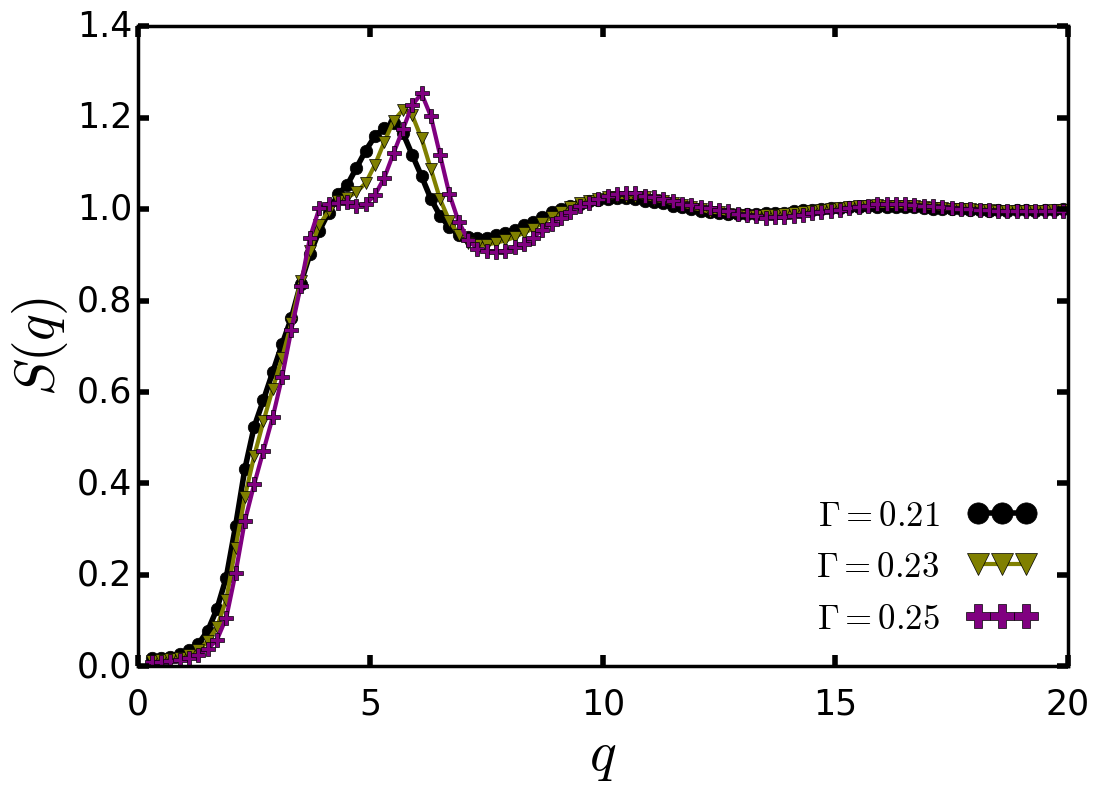}
         \caption{}
     \end{subfigure}
     \hfill
     \begin{subfigure}[b]{0.45\textwidth}
         \centering
         \includegraphics[width=\textwidth]
         {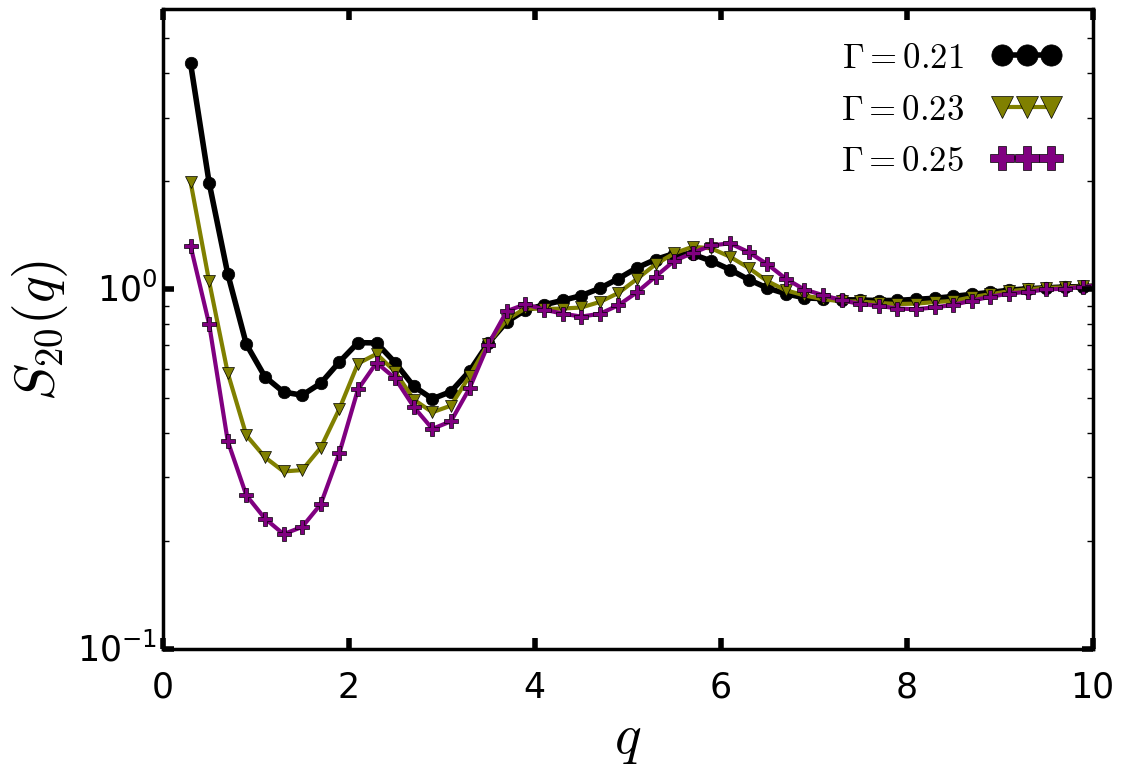}
         \caption{}
     \end{subfigure}
     \hfill
        \caption{The pair correlation function $g(r)$, the structure factor $S(q)$, and the orientation structure factor $S_{20}(q)$ for different nematic states in the simulation; see legend. The x-axis is not scaled here.}
        \label{rdfSqSq20_nem}
\end{figure}
In the simulations, soft ellipsoids with aspect ratio $(a/b)=3.5$ are found to form isotropic states at densities $\Gamma < 0.21$. This can be inferred from the plots of the pair correlation function $g(r)$ and the structure factor $S(q)$ in Fig.~\ref{rdfSq_isotropics} where there are no long range structures. In addition, the isotropic states manifest no nematic order which can be noticed from the decay of the orientation correlation in $G_2(r)$ to zero across the system length scales as noticed in Fig.~\ref{ordfIsoNem}. To inspect that there is no nematic order during the simulation time, the scalar nematic order parameter $S$ is measured and shows a very weak and negligible nematic order (not shown). We attribute it to the finite simulation box size.  In the plots of $g(r)$ and $S(q)$ we show comparisons between the soft ellipsoids at different densities $\Gamma$ and the hard ellipsoids as experimentally measured at different volume fractions $\phi_{exp}$. We define a ratio of length-scales as mapping parameter $\lambda={\sigma^{H}_0}/\Delta$ (where $\Delta$ is some length scale) so that we can \textit{map} the soft-ellipsoid structure \textit{onto} the hard ellipsoid structure. Basically, we scale the x-axis of the soft-ellipsoid plots of the $g(r)$ and $S(q)$ functions by the factor $\lambda$ until the structure of soft ellipsoids matches the structure of the hard ellipsoids. This means that for the structure plots of the hard ellipsoids we have $\lambda={\sigma^{H}_0}/{\sigma^H_0}=1$, with $\sigma_0^H=2\times 1.23\;\mu m$ being the small diameter of the hard ellipsoid, since the hard ellipsoid plots are our reference plots. For the \textit{scaled} soft ellipsoid plots, the best mapping parameter for the isotropic states is found to be $\lambda={\sigma^{H}_0}/{\sigma^{S}_0}=0.83$. Considering the translation degrees of freedoms, it is obvious from the $g(r)$ and $S(q)$ plots that the system structures in the simulation for the isotropic densities $\Gamma=0.16$ and $\Gamma=0.18$ are best mapped to the system structures in the experiment for the isotropic volume fractions $\phi=0.40$ and $\phi=0.46$, respectively.

At this point, the spatial orientation structure of the isotropic states needs to be discussed. The upper panel in Fig.~\ref{ordfoSq_isotropics} presents the orientation radial distribution function $G_2(r)$ in an isotropic state.  It is clear from the plot that the orientation correlations in the simulation, carried out with the anisotropic RGB potential, are more enhanced than the ones in the hard-ellipsoid experiment. Additionally, the $G_2(r)$ from EDBD simulations of hard ellipsoids \cite{RollerLaganapan2020} is included in Fig.~\ref{ordfoSq_isotropics}. Its lower values  reflect the fact that the  soft ellipsoid neighbours align more strongly  than the hard ones. This demonstrates the effect of the particle softness in the simulation which prevents producing unstable forces and torques at small length scales. The difference between the local angular structure in the soft-ellipsoid simulation and the hard-ellipsoid experiment results from the fact that we use an anisotropic potential in the simulation which enhances the local alignments of the soft ellipsoids. The orientation structure factor $S_{20}(q)$ is shown in the lower panel of Fig.~\ref{ordfoSq_isotropics}. The large wave vector oscillations show the discussed local differences, but the long-ranged orientation correlations become stronger on increasing the density equally in simulation and experiment.

When increasing the densities to larger than $\Gamma \ge 0.21$, the soft-ellipsoid system transforms from the isotropic to the nematic phase. As there are no experimental data available for the nematic phase, we show the results of this phase only in the simulation. What characterizes the nematic states are the finite limit of the function $G_2(r)$ at far distances and its clear peaks as seen in Fig.~\ref{ordfIsoNem}. Additionally, the values of the nematic scalar order parameter $S$ 
explicitly prove the existence of the nematic phase for $\Gamma \ge 0.21$. {They will be  obtained from a finite size analysis performed in Fig.~\ref{clusterAnalysis}; see there for more details.} It is clear that when increasing the nematic density more and more, the degree of the nematic order increases as can be seen in the increasing values of $G_2(r\gg 1)$ and $S$, that fulfill \cite{Eppenga1984} $G_2(r\to\infty ) = S^2 $. 
Probing the structures in the plots of the functions $g(r)$ and $S(q)$ in Fig.~\ref{rdfSqSq20_nem}, we observe that the nematic states have more translation order than the isotropic states. The translation order is reinforced when the system density increases. The presence of a small peak in the $S(q)$ plot around $q\approx4.0$ in Fig.~\ref{rdfSqSq20_nem} reflects the emergence of a weak translation order at that length scale. This weak translation order results from the strong orientation order at the short length scales in the orientation structure factor of the nematic states $S_{20}(q)$ around $q\approx2.3$ and $q\approx3.8$ in the same figure. Once again, the softness of the particles plays an important role in having such short and stable orders in the nematic states. An interesting feature of the function $S_{20}$ is that it informs about the long-wavelength orientation density fluctuations when looking at its small wave vector regime. It is clear that at small wave vectors in Fig.~\ref{rdfSqSq20_nem} the value of $S_{20}$ decreases when the fluid density increases. This is due to the fact that the amplitudes of the local orientation density fluctuations become smaller when the system moves away from the density at which the isotropic-nematic phase transition occurs ($\Gamma\approx 0.21$). In the isotropic phase, the same phenomenon happens but in the opposite direction. When the system density approaches the isotropic-nematic transition, the value of $S_{20}$ at small wave vectors increases which is a sign that the magnitude of orientation density fluctuations increases so that these fluctuations dominate the whole simulation box at the phase transition.

\begin{figure}[h!]
\centering
\includegraphics[width=0.45\textwidth]{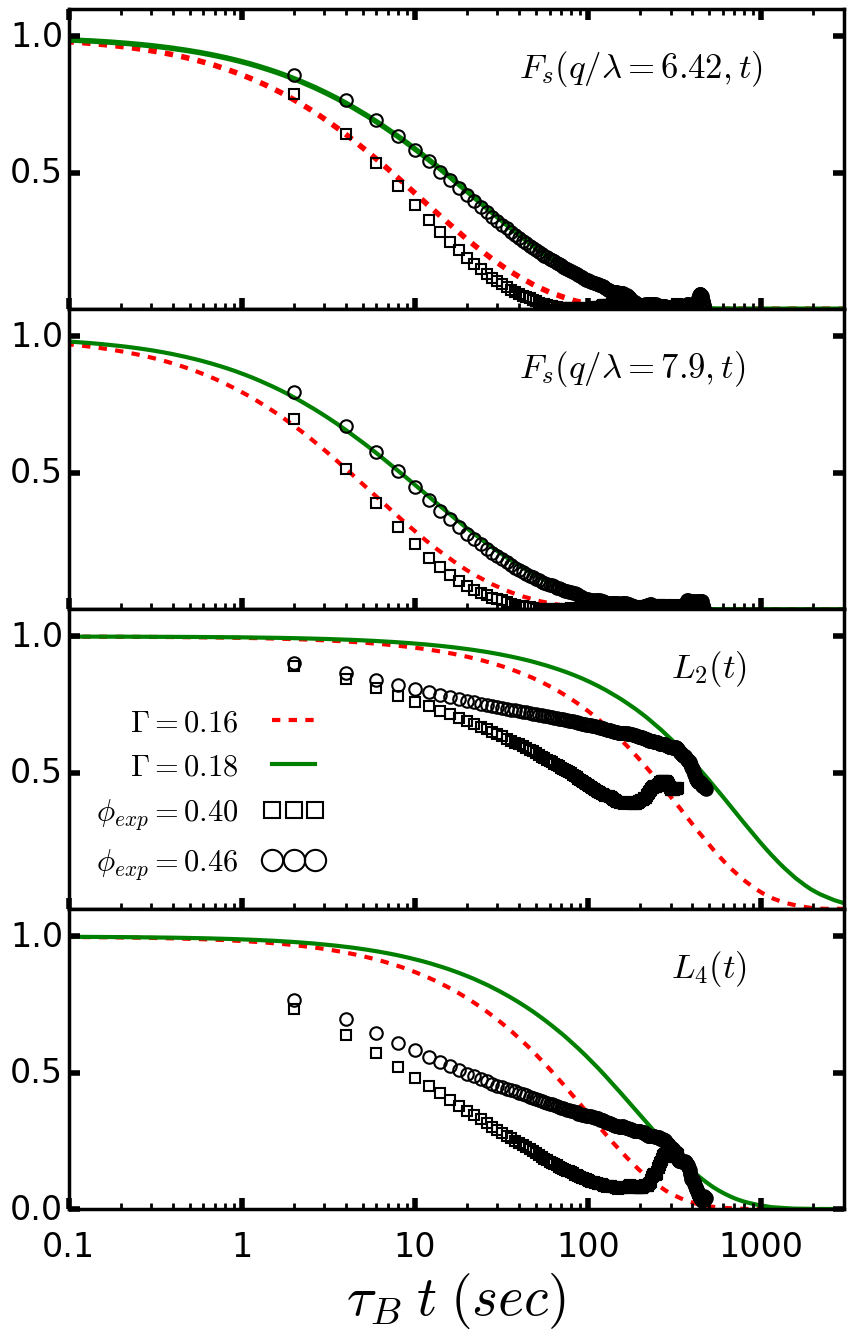}
\caption{Dynamics of the isotropic states in the simulation and experiment. For $\Gamma=0.16$, $\tau_B=140 \;sec$. For $\Gamma=0.18$, $\tau_B=220 \;sec$. The experiment (hard ellipsoid) data are the squares and the circles. The solid and the dashed lines are the simulation (soft ellipsoid) curves. }
\label{sisf_ocf_iso}
\end{figure}

Having discussed the structures of the isotropic and nematic states, we now discus their dynamics and show the results of the translation and orientation correlation functions in both equilibrium phases. Figure \ref{sisf_ocf_iso} shows the self intermediate scattering function $F_s(q/\lambda,t)$ and the orientation correlation functions $L_2(t)$ and $L_4(t)$ in the isotropic states for the soft ellipsoids (simulation) and the hard ellipsoids (experiment) fluids. Notice that the time axis is in seconds,  and that a single Brownian time $\tau_B$ depending on the density via hydrodynamic interactions matches the dynamics of all time-dependent functions. The factor $\lambda$ is present in the $F_s$ plots in order to have the correct mapping between the soft and hard ellipsoids at the respective length scales. It is obvious from Fig.~\ref{sisf_ocf_iso} that the relaxation of the translation dynamics in the simulation agrees almost exactly with that in the experiment. This can be seen in the plots of the self intermediate scattering function $F_s$ for the two wave vectors $(q/\lambda)=6.42$ and $(q/\lambda)=7.9$. We repeat here that by setting $\lambda=1$ we obtain the hard-ellipsoid value, while setting $\lambda=0.83$ we obtain the \textit{scaled} soft-ellipsoid value. The first vector $(q/\lambda)=6.42$ corresponds to the peak in the isotropic structure factor and the second $(q/\lambda)=7.9$ is a vector close to the peak seen in Fig.~\ref{rdfSq_isotropics}. Observing the decay of the self intermediate scattering function in the simulation and the experiment in Fig.~\ref{sisf_ocf_iso}, we deduce that the dynamics of the states $\Gamma=0.16$ and $\Gamma=0.18$ agree with the dynamics of the states $\phi_{exp}=0.40$ and $\phi_{exp}=0.46$ respectively. These agreements in the isotropic-phase dynamics between simulation and experiment have also been seen in the isotropic-phase structure functions $g(r)$ and $S(q)$ in Fig.~\ref{rdfSq_isotropics}. Also, the relaxations of the rotation degrees of freedoms depicted by the functions $L_2$ and $L_4$ in Fig.~\ref{sisf_ocf_iso} show agreements and discrepancies in the rotation dynamics between the soft ellipsoids and the hard ellipsoids. In both systems, the soft ellipsoids and the hard ellipsoids, the rotation correlations seen in $L_2$ and $L_4$ tend to live longer and decay on rotation timescales larger than the translation timescales at which the $F_s$ functions decay. {The rotation timescales} are larger than the translation timescales by one to two decades in both the soft and hard ellipsoid systems. At the same time, the decay of the rotation dynamics in the hard ellipsoids sets in at earlier times than the one in the soft ellipsoids. An explanation for the discrepancy in the rotation correlation functions between simulation and experiment is the inevitable noise in the experiment that makes the correlation functions in the experiment decorrelate earlier than those of the simulation.
\begin{figure}[h!]
 \centering
 \includegraphics[width=0.45\textwidth]{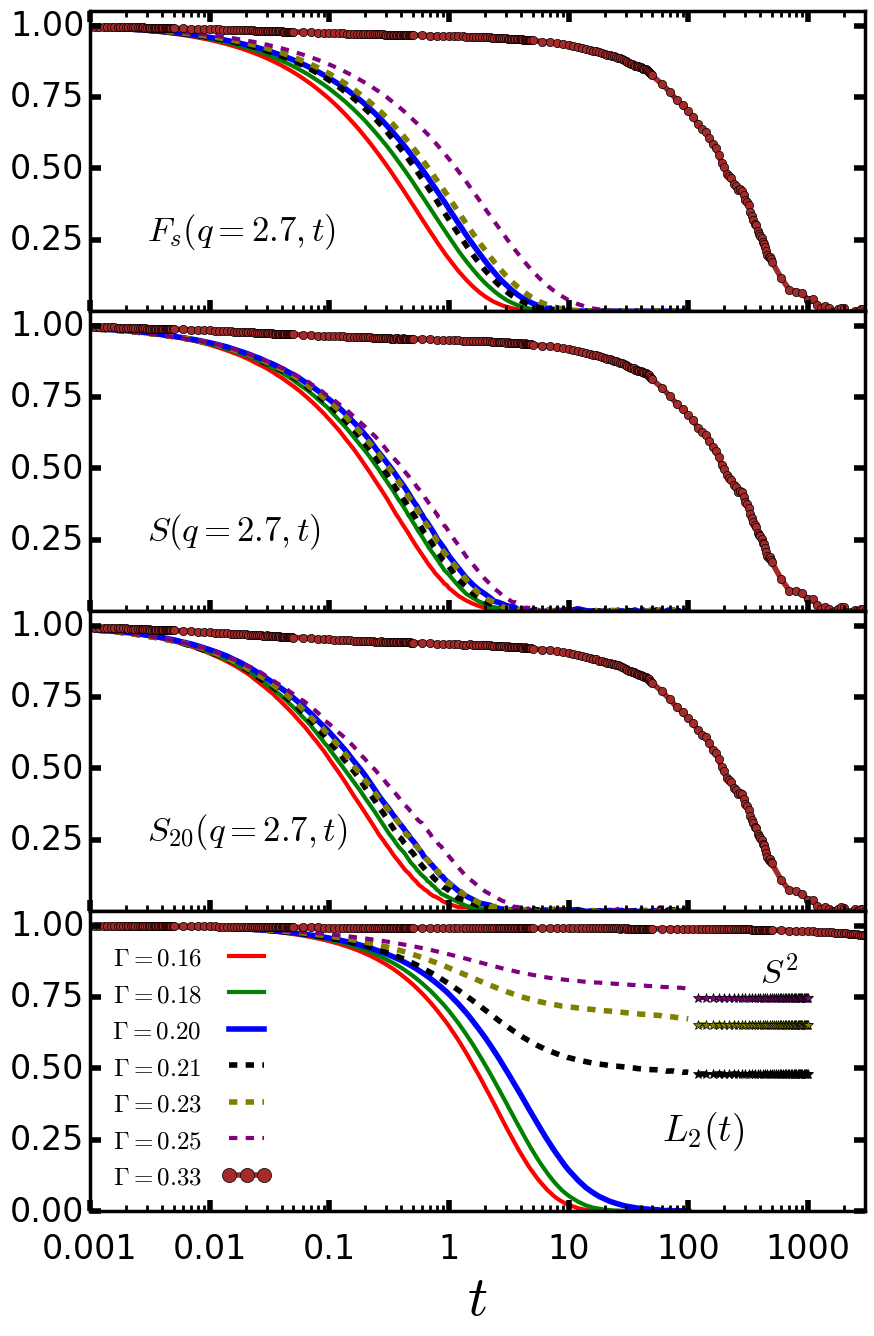}
 \caption{Dynamics of the isotropic, nematic, and liquid glass states in the simulation. See the text for more details. The isotropic states are the solid lines. The nematic states are the dashed lines {with $S^2$ added from Fig.~\ref{clusterAnalysis} (at particle number $\approx 275$)}. The liquid glass state $\Gamma=0.33$ is the solid line marked with the filled circles.}
 \label{sisf_ocf_iso_nem_lg}
\end{figure}

Now, we concentrate our discussion only on the simulation results and show an overview of the dynamics in the isotropic and nematic states, as well as in the state of the liquid glass.
Figure \ref{sisf_ocf_iso_nem_lg} depicts the plots of the translation correlation functions $F_s(q,t)$, $S(q,t)$, and $S_{20}(q,t)$ at the wave vector $q=2.7$, and the orientation correlation function $L_2(t)$. Notice the absence of the factor $\lambda$ which means that the simulation plots are not rescaled because there are no experimental (hard ellipsoids) data. The wave vector $q=2.7$ is close to the first small peak in the plot of $S_{20}(q)$ of the nematic states in Fig.~\ref{rdfSqSq20_nem}. Figure \ref{sisf_ocf_iso_nem_lg} shows the isotropic states ($\Gamma < 0.21$ with solid lines), the nematic states ($\Gamma \ge 0.21$ with dashed lines), and a liquid glass state ($\Gamma=0.33$ with a solid line marked by circles). The fastest decay in the isotropic states corresponds to $\Gamma=0.16$ while the slowest corresponds to $\Gamma=0.20$. The fastest decay in the nematic states corresponds to $\Gamma=0.21$ while the slowest corresponds to $\Gamma=0.25$. In the \hyperref[s:suppMater]{supplementary materials}, 
results are given for the structure and dynamics of the additional states $\Gamma=0.27, 0.29, 0.31$. As expected, in each phase the relaxation timescale of the correlation functions increases when the density increases. At each density, the rotation timescale seen in $L_2(t)$ in Fig.~\ref{sisf_ocf_iso_nem_lg} is remarkably larger than the translation timescales observed in $F_s(q,t)$, $S(q,t)$, and $S_{20}(q,t)$ for wave vectors connected to the structural relaxation. The simulation time for the isotropic and nematic states stops once the states are equilibrated (at $t=100$). We had to run much longer simulations for the state $\Gamma=0.33$, and yet could not make angular correlations reach equilibrium. The simulation time for the state $\Gamma=0.33$ ends at $t=3000$ (see Fig.~\ref{sisf_ocf_iso_nem_lg}). In the nematic phase, the $L_2(t)$ function decays to a nonzero value which is the quadratic value of the equilibrium scalar nematic order parameter. Furthermore, the decays of the functions $F_s(q,t)$, $S(q,t)$, and $S_{20}(q,t)$ in Fig.~\ref{sisf_ocf_iso_nem_lg} for $\Gamma=0.21$ are faster than the decays of the same functions at the isotropic state $\Gamma=0.20$. This reflect the fact that when the particles cross the isotropic-nematic transition, their centers of masses can move more freely due to the orientation order which was absent in the isotropic state $\Gamma=0.20$. However, this is only noticed close to the isotropic-nematic transition. At $\Gamma=0.33$, a liquid glass state can be identified. From Fig.~\ref{sisf_ocf_iso_nem_lg}, the most obvious dynamical behavior of the liquid glass is the decay of the translation correlations in $F_s(q,t)$, $S(q,t)$, and $S_{20}(q,t)$ around timescales $t\approx 700$ while the orientation correlation in $L_2(t)$ persists with a very high amplitude. This reflects the fact that the particle rotation dynamics are arrested while the ellipsoid centers of masses are moving freely which enables the translation correlations to equilibrate. While in the nematic state, the arrest in the dynamic orientation correlations is in agreement with the measured nematic order, in the liquid glass, where $S$ is very small (see Fig.~\ref{clusterAnalysis}), this relation clearly does not hold.

\subsection{The Liquid Glass}
\label{ss:liquid glass}
\begin{figure}[h!]
     \centering
     \begin{subfigure}[b]{0.23\textwidth}
         \centering
         \includegraphics[width=\textwidth]
         {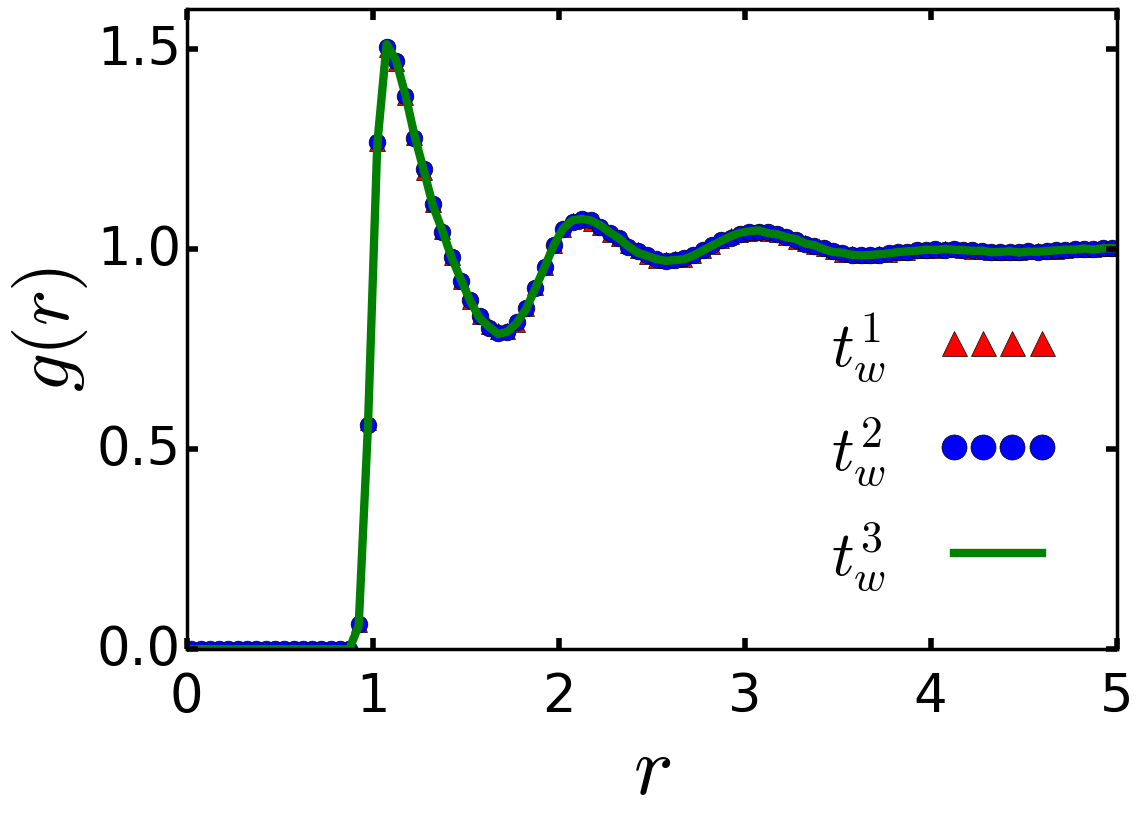}
         \caption{}
     \end{subfigure}
     \hfill
     \begin{subfigure}[b]{0.23\textwidth}
         \centering
         \includegraphics[width=\textwidth]
         {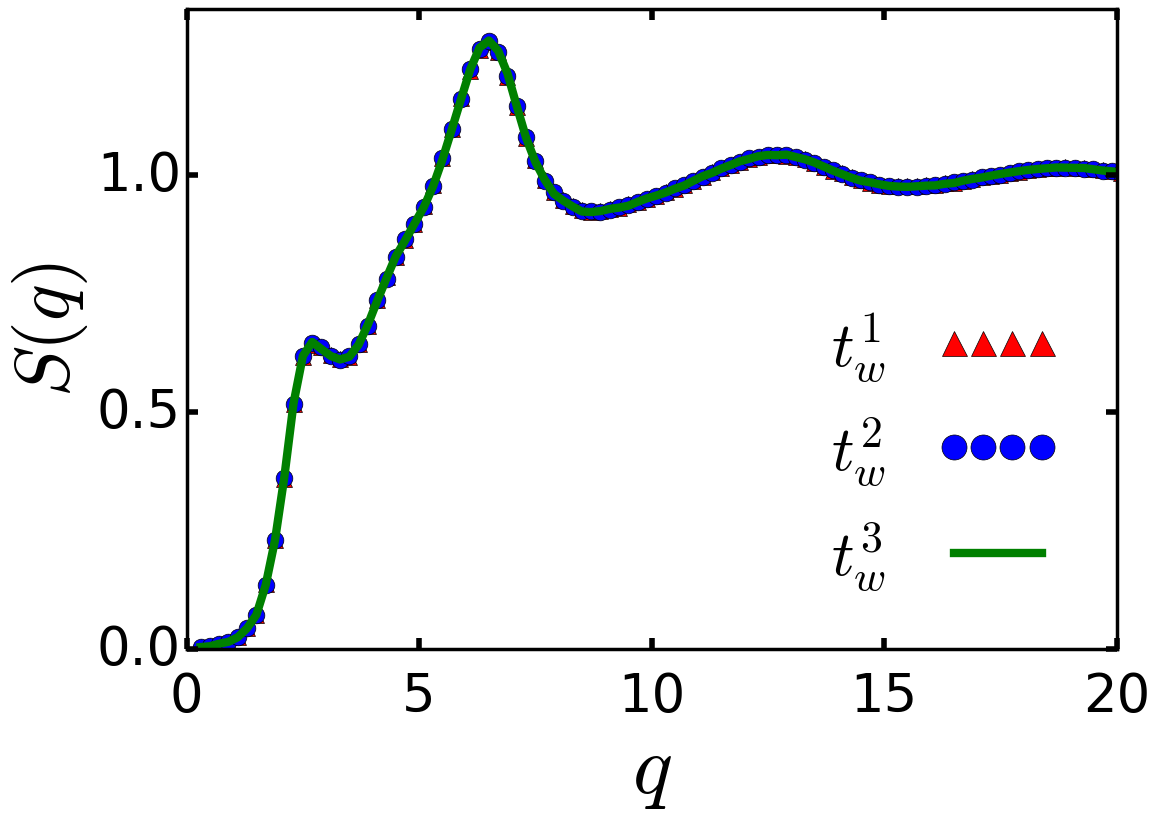}
         \caption{}
     \end{subfigure}
     \hfill
     \begin{subfigure}[b]{0.23\textwidth}
         \centering
         \includegraphics[width=\textwidth]
         {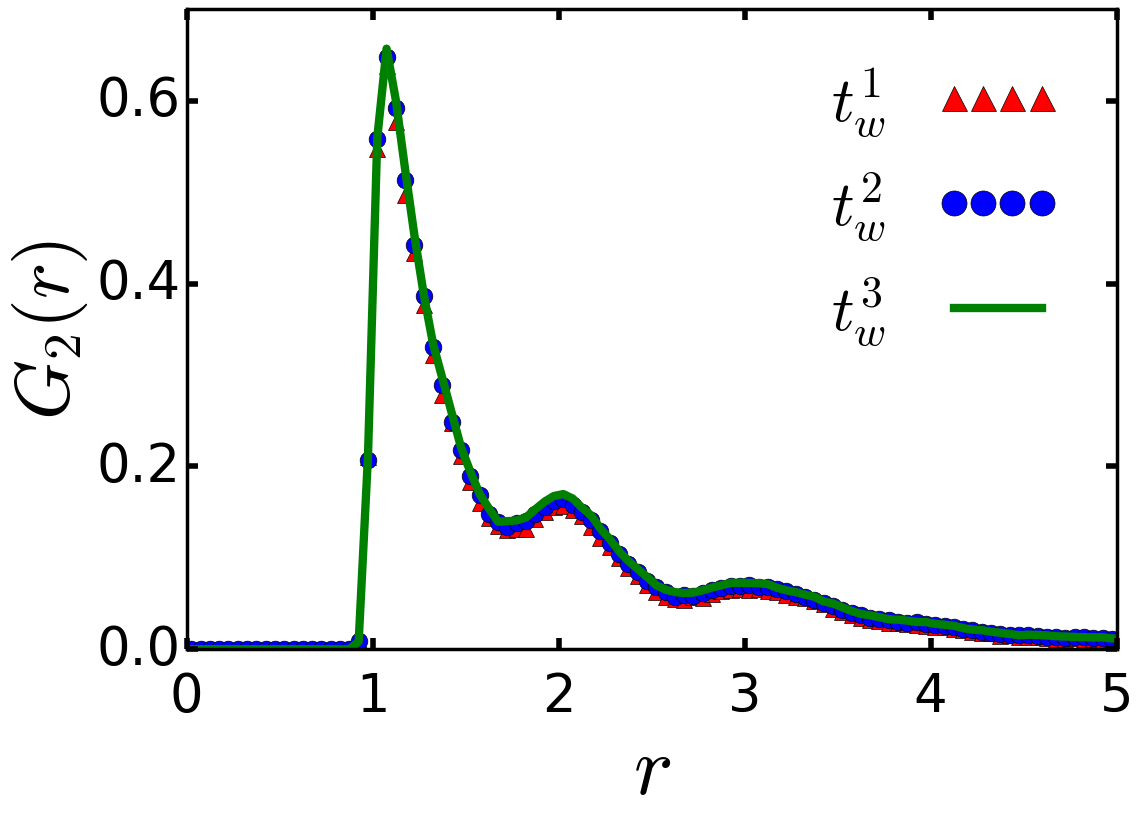}
         \caption{}
     \end{subfigure}
     \hfill
     \begin{subfigure}[b]{0.23\textwidth}
         \centering
         \includegraphics[width=\textwidth]
         {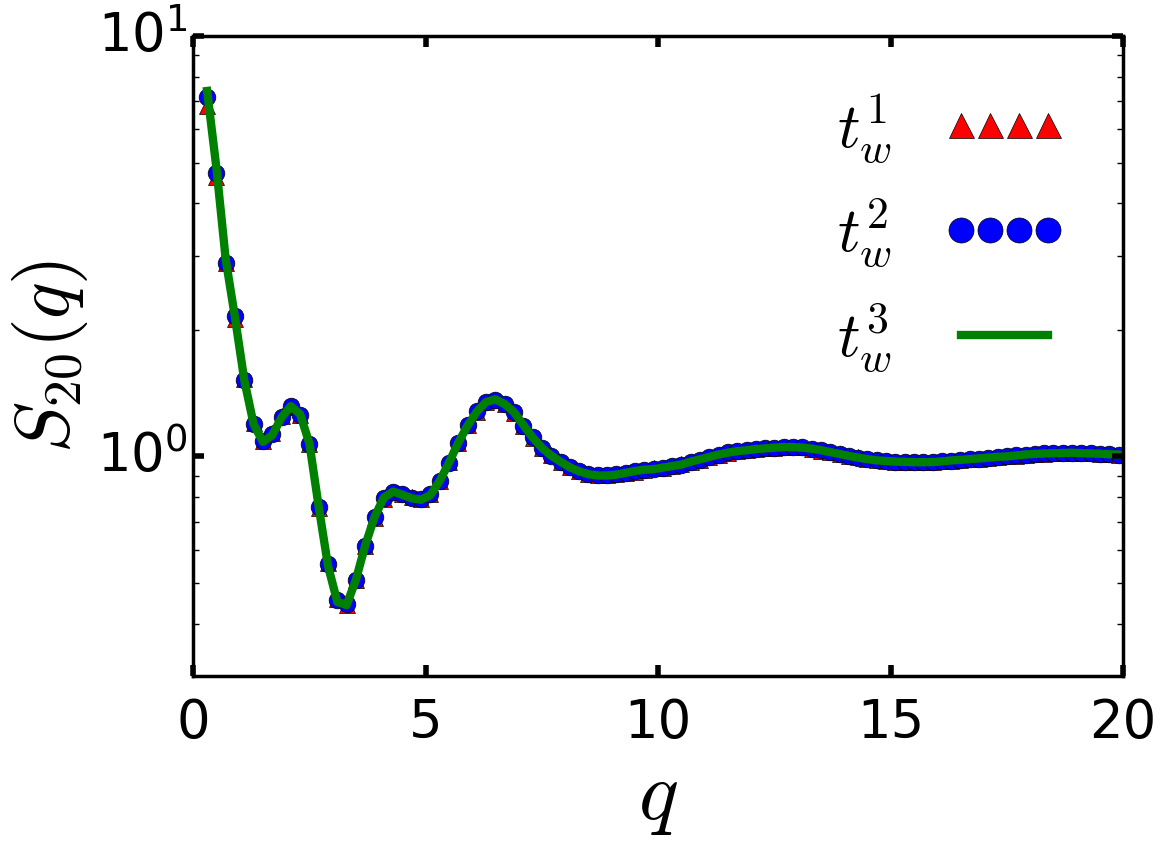}
         \caption{}
     \end{subfigure}
     \hfill
        \caption{Structure functions sensitive to packing and alignment, $g(r)$, $S(q)$, $G_2(r)$, and $S_{20}(q)$, in the simulation at density $\Gamma=0.33$ for three different simulations, i.e., three different waiting times $t^1_w$, $t^2_w$ and $t^3_w$ with $t^i_w=3000\times i$ where $i \in {1,2,3}$. The decay of $G_2(r)$ to zero as $r$ increases is a strong indication of the isotropic orientation correlations.}
        \label{rdf_m2m3m4_33}
\end{figure}
We now focus on the measurement phase of the liquid glass states at the density $\Gamma=0.33$. The success in quenching into such a high density heavily relies on the nature of the particles used in the simulation. Introducing the particle softness by using the RGB potential helps to avoid the formations of unstable physical configurations upon quenching. Such softness does not generate any large forces that cease the MD simulation. 
\begin{figure}[h!]
     \centering
     \begin{subfigure}[b]{0.45\textwidth}
         \centering
         \includegraphics[width=\textwidth]
         {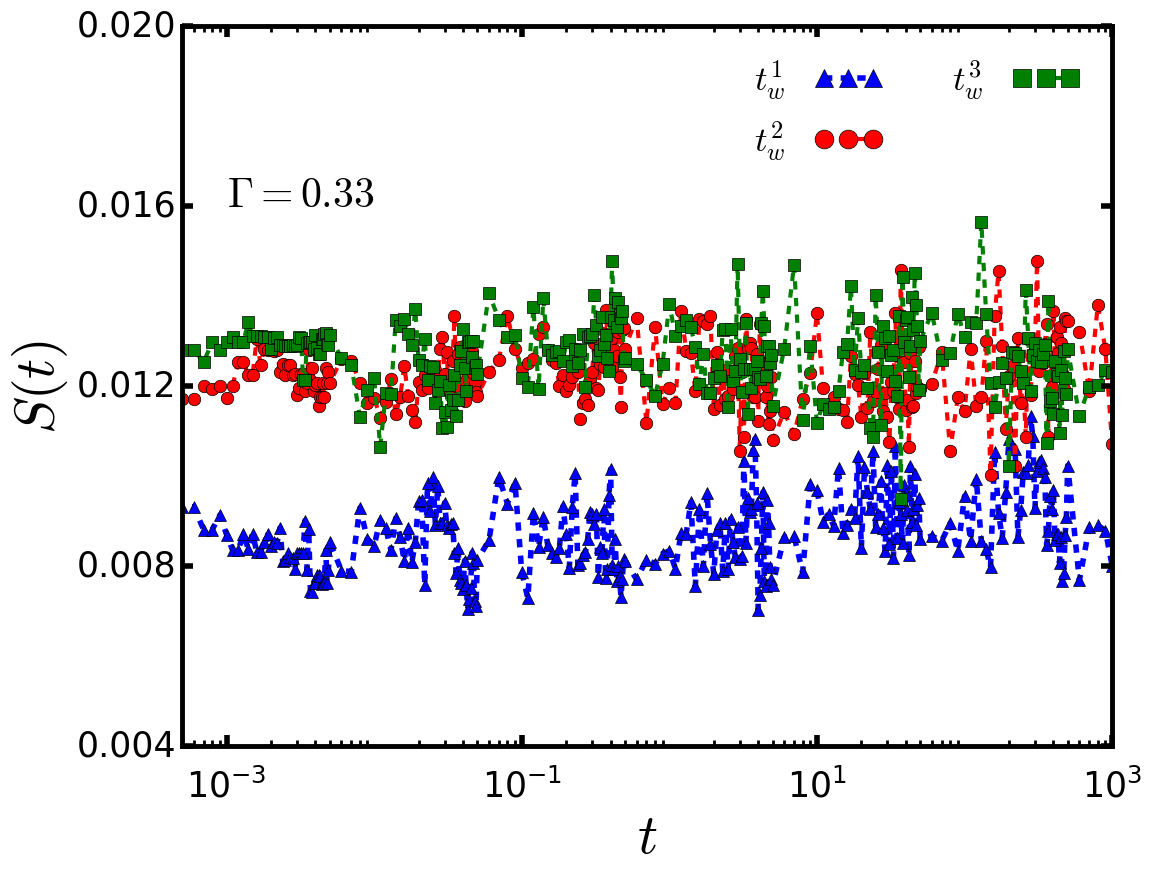}
         \caption{}
     \end{subfigure}
     \hfill
     \begin{subfigure}[b]{0.45\textwidth}
         \centering
         \includegraphics[width=\textwidth]
         {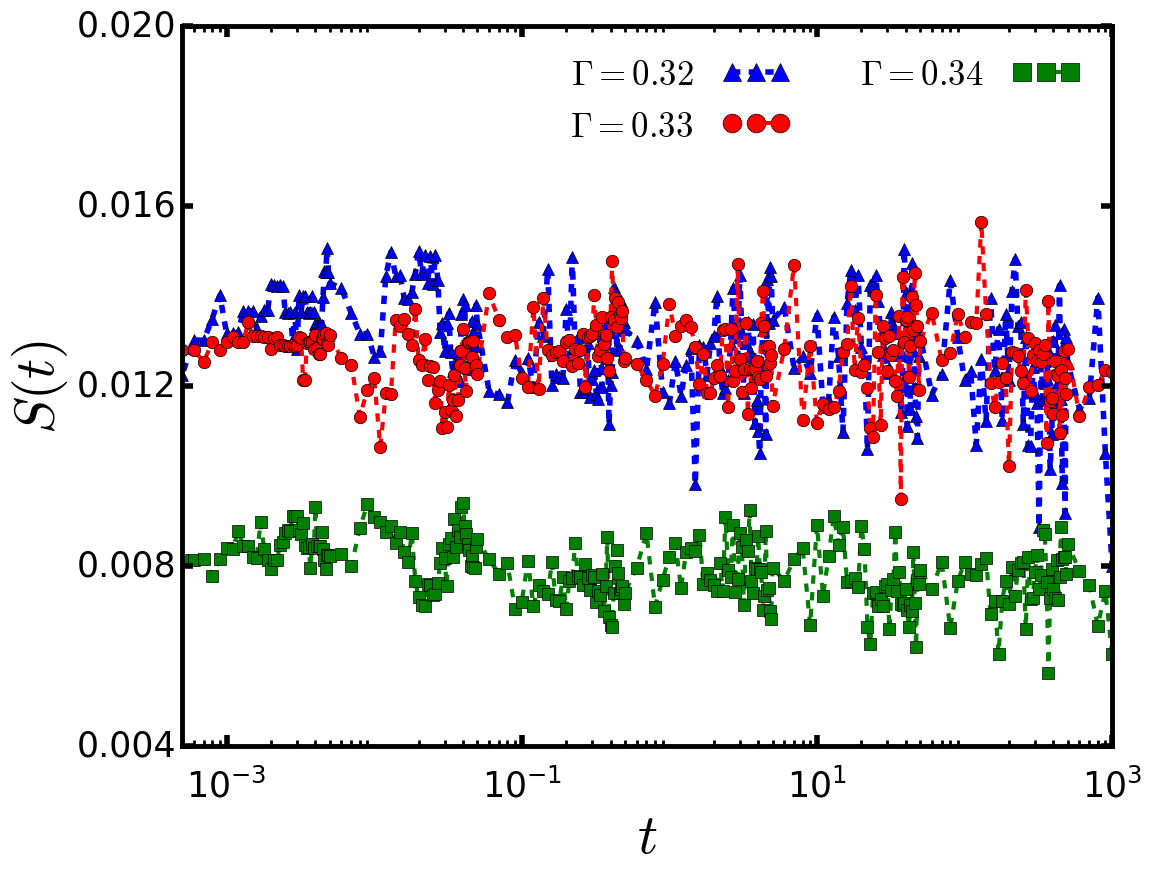}
         \caption{}
     \end{subfigure}
     \hfill
        \caption{Simulation: The scalar nematic order parameter $S(t)$ for different waiting times for $\Gamma=0.33$ (upper panel) and for different densities for the waiting time $t^3_w$ (lower panel). It is clear that the liquid glass does not show any significant nematic order.}
        \label{diagQ_lg_m2m3m4_diffGammas}
\end{figure}

\begin{figure}[h!]
     \centering
     \includegraphics[width=0.45\textwidth]{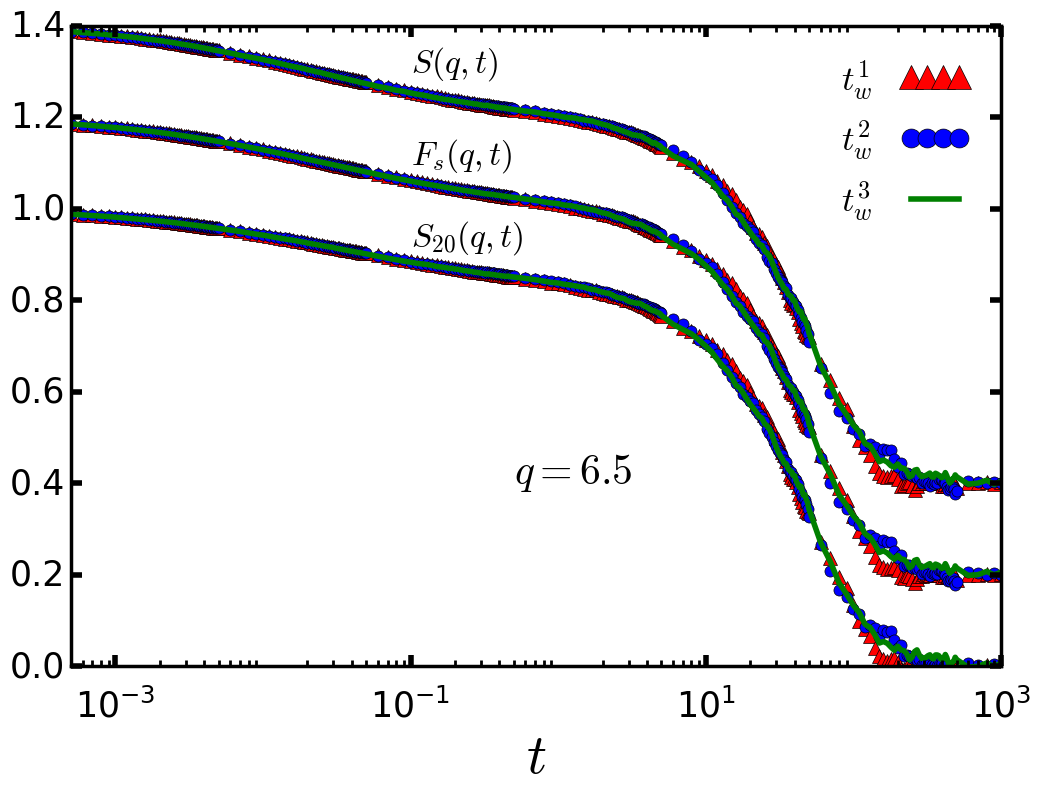}
     \hfill
     \caption{Translation correlation functions in the simulation for $\Gamma=0.33$ at $q=6.5$. $F_s(q,t)$ and $S(q,t)$ are shifted vertically for a better view. The curves at three different waiting times $t^i_w$ show that the translation dynamics do not manifest any aging; the translation degrees of freedom relax like in a fluid.}
        \label{transCorr_m2m3m4_33}
\end{figure}

\begin{figure}[h!]
     \centering
     \includegraphics[width=0.45\textwidth]{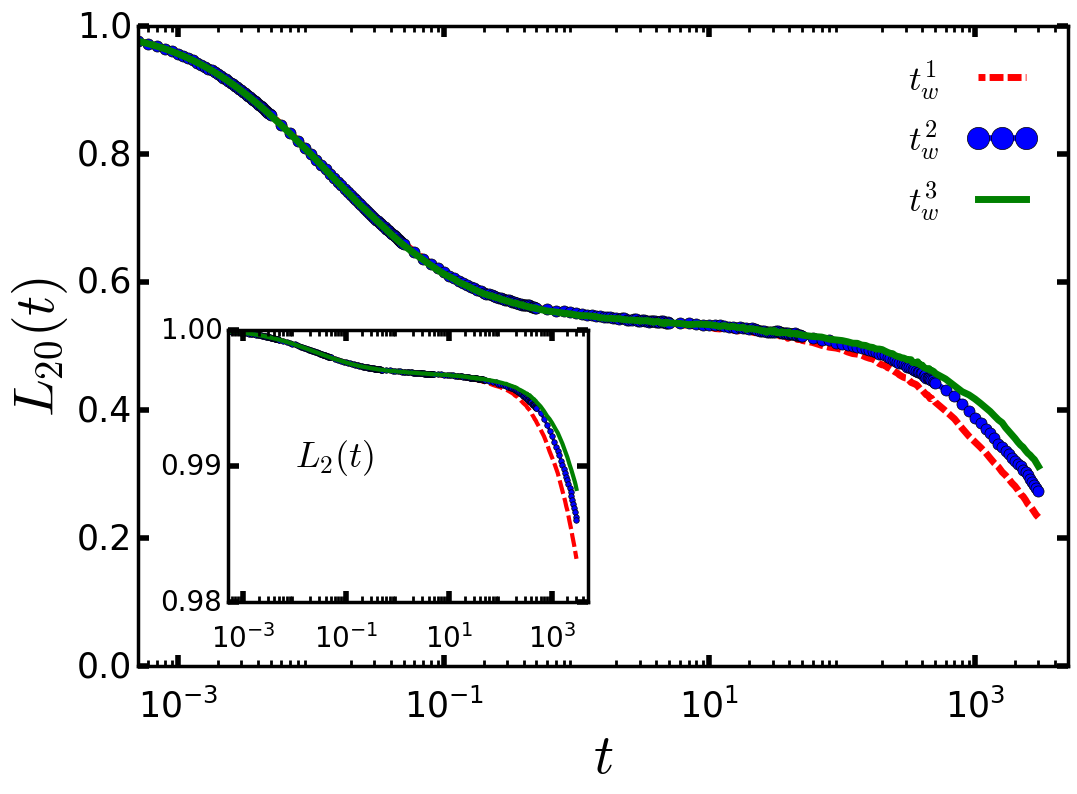}
     \hfill
        \caption{The orientation correlation functions $L_{20}(t)$ in the simulation at density $\Gamma=0.33$ for three different waiting times $t^i_w$ as labeled. Aging is observed by the dependence of the final decay on $t^i_w$. The aging seen in the three simulations evidences that the orientation motion can not relax in our simulation time window. The inset shows $L_2(t)$ which requires larger scale angular motion to decorrelate; thus aging arises in the first 2 \% of the relaxation.}
        \label{orientCorr_m2m3m4_33}
\end{figure}

At the liquid glass densities, the anisotropic system becomes arrested in a metastable state identified by a plateau in the translation correlation functions $\phi^{t}_q(t)$ such as $F_s(q,t)$ and another plateau in the rotation correlation functions $\phi^r_n(t)$ such as $L_n(t)$. The possibility that at such high densities the system might develop  \textit{long} range translation and orientation orders at the same time, viz.~enter the crystalline phase, can be inspected by looking at the structure functions in Fig.~\ref{rdf_m2m3m4_33}. The meanings of $t^1_w$, $t^2_w$ and $t^3_w$ in the figure refer to the waiting  times for every plot to be produced in the simulation with $t^i_w=3000\times i$ where $i \in {1,2,3}$. {This convention is adopted for {Figs. \ref{rdf_m2m3m4_33} to \ref{orientCorr_m2m3m4_33}}}. Looking at the curves in Fig.~\ref{rdf_m2m3m4_33}, one can observe the lack of long range translation and orientation orders. In particular, $G_2(r)$ goes to zero as $r$ increases which is a sign that the degree of alignment in the system at $\Gamma=0.33$ is neither nematic nor crystalline \cite{Rodriguez-Odriozola, AdamsLuckhurstPhippen1987}: Absence of crystallinity is obvious from the plots of $g(r)$, $G_2(r)$, $S(q)$ and $S_{20}(q)$ in Fig.~\ref{rdf_m2m3m4_33} in the three different simulations (three different waiting times). An interesting feature at small wave vectors is a small peak in $S(q)$ at $q \approx 2.7$ which results from the emergence of the orientation correlations characterized by two distinct peaks in $S_{20}(q)$ at $q \approx 2.1$ and $q \approx 4.2$. This is reminiscent of the short range orientation order in the plot of $S_{20}(q)$ of the nematic phase in Fig.~\ref{rdfSqSq20_nem}. The two peaks indicate that the development of the nematic phase is hindered by the slowing-down of the glassy dynamics. The presence of the two peaks has not been reported before, and it sheds light onto a strong local orientation order in the liquid glass at small wave vectors. We believe that the enhanced local angular structure results from anisotropic interactions. Contrary to the nematic states, the high density in the liquid glass state suppresses the free alignment of the ellipsoids and no substantial nematic order forms over the length scales in the simulation box. The scalar nematic order parameter $S(t)$ for the three waiting times is shown in Fig.~\ref{diagQ_lg_m2m3m4_diffGammas}. It is obvious that there is no substantial nematic order in the three simulations. Probing the structure of the glassy state only does not enable us to identify the glass state. From Fig.~\ref{rdf_m2m3m4_33} one cannot confidently distinguish between the structure of such a glassy state and the structure of the isotropic fluid discussed above. This aspect is typical of glassy systems. Hence the dynamics of the state $\Gamma=0.33$ need to be discussed to obtain a conclusive explanation.

Concerning liquid glass dynamics, first the translation correlation functions shall be discussed. Fig.~\ref{transCorr_m2m3m4_33} shows the plots of the dynamic structure factor $S(q,t)$, the self intermediate scattering function $F_s(q,t)$, and the dynamic orientation structure factor $S_{20}(q,t)$. These functions are computed for the wave vector of the main peak in the structure factor $S(q=6.5)$ as it can be seen in Fig.~\ref{rdf_m2m3m4_33}. It is clear that these functions do not show aging in the three simulations, viz.~for the three waiting times. Interestingly, the translation dynamics illustrated in Fig.~\ref{transCorr_m2m3m4_33} behaves as the dynamics characterizing the supercooled hard-sphere fluid. {After the initial diffusive regime, the  dynamics  relaxes slowly onto a fairly flat plateau and then decays  below it. This functional form} can be seen in the three translation correlation functions. The plateau is a clear signal of the existence of a cage. The final decay  takes the form of a stretched-exponential curve with an initial power-law decay. It describes  that the particles leave the cages which impede their translation motion for some intermediate time; see Sect.~\ref{s:MCT} for more detail. 

The orientation correlation functions exhibit a different type of dynamics. Figure \ref{orientCorr_m2m3m4_33} depicts how the orientation correlations evolve in the simulation. The figure shows $L_{20}(t)$. Because of their high plateaus, low orders of $L_n(t)$ would not give such clear indications for two step relaxation. As an example the plot of $L_2(t)$ is shown in the inset of Fig.~\ref{orientCorr_m2m3m4_33}. The reason for this is the tight arrest of the rotation motion of the ellipsoids. Hence, a higher order rotation correlation function ($n=20$) is required in order to resolve it. It is obvious from the figure that the ellipsoid rotation dynamics decay from the initial diffusive regime onto a plateau due to an angular cage which arrests the rotation motion. Later, the system dynamics fail to follow the $\alpha$ relaxation decay. The aging phenomenon causes the particle rotation dynamics to relax which is different from the case of the translation dynamics. The longer the simulation, the longer the plateau lives, which means that the particle orientations are arrested for longer times. The aging indicates that the system always finds a new metastable liquid glass state which has a deeper minimum in the free energy landscape of the system. While the orientation correlations are not able to relax, the translation dynamics has already relaxed much earlier around $t\approx 50$ (see figures \ref{transCorr_m2m3m4_33} and \ref{orientCorr_m2m3m4_33}). The lifetime of the translation plateau is around two decades (and it is followed by a final relaxation), whereas the lifetime of the orientation plateau is around three decades. Hence, the observed type of cage is an \textit{orientation cage} which allows the translation degrees of freedom to decorrelate but arrests the orientation ones. This confirms an assignment of the state as a liquid glass since the ellipsoids can move while their orientations remain arrested. The height of the plateau and the tightness of the cage will quantitatively be discussed in the framework of mode coupling theory for the glass transition in section \ref{s:MCT}.
\begin{figure}[h!]
     \centering
     \begin{subfigure}[b]{0.45\textwidth}
         \centering
         \includegraphics[width=\textwidth]
         {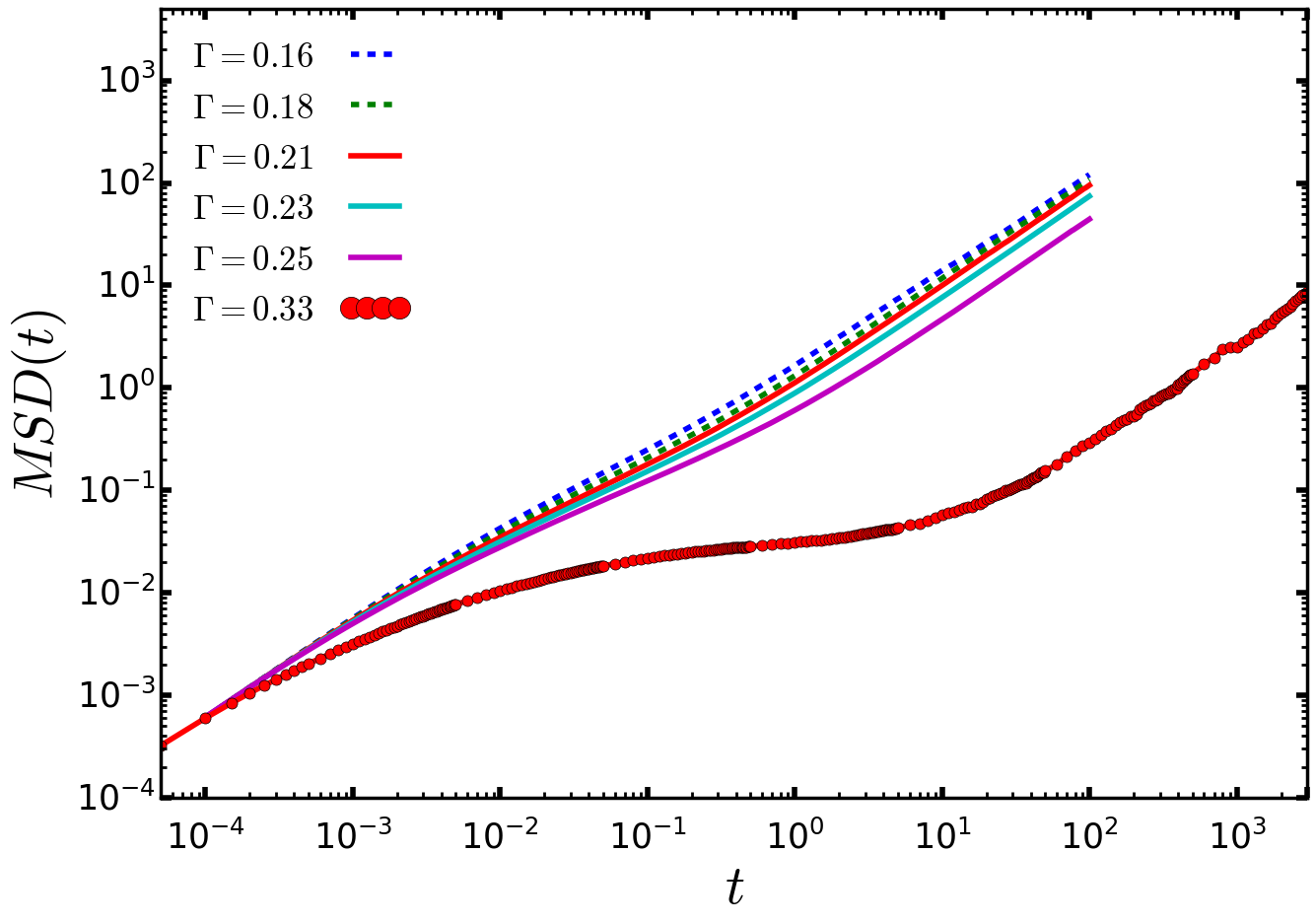}
         \caption{}
     \end{subfigure}
     \hfill
     \begin{subfigure}[b]{0.45\textwidth}
         \centering
         \includegraphics[width=\textwidth]
         {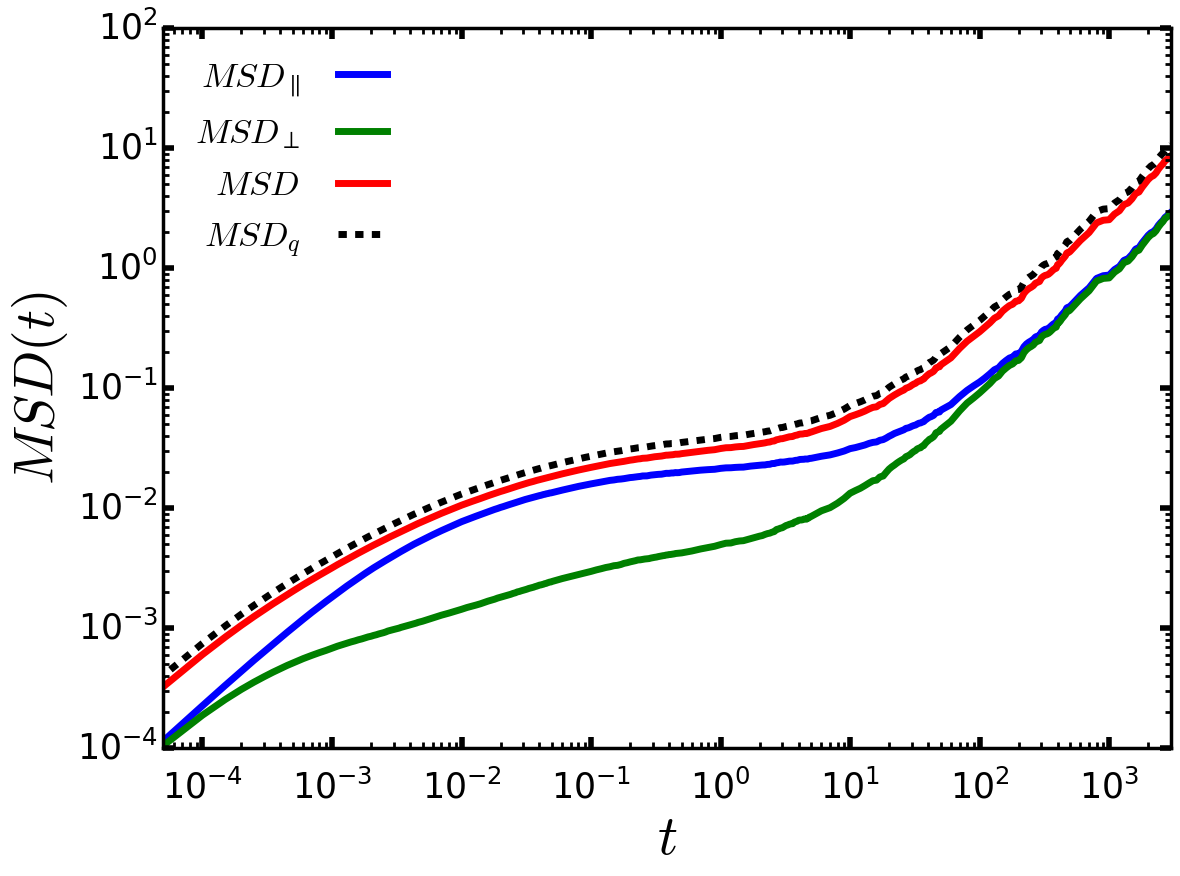}
         \caption{}
     \end{subfigure}
     \hfill
        \caption{Top: Mean squared displacements (MSD) for the isotropic ($\Gamma=0.16, 0.18$), the nematic ($\Gamma=0.21 - 0.25$), and the liquid glass ($\Gamma= 0.33$) state. Bottom: The MSD for the liquid glass state $\Gamma=0.33$ as well as its parallel (MSD$_{\parallel}$) and perpendicular (MSD$_{\perp}$) components. The plot of MSD$_q$ is obtained by using the self intermediate scattering function at $q=0.3$ from the Gaussian approximation MSD$_q=(-6/q^2)\ln(F_s(q=0.3,t))$. }
        \label{msd}
\end{figure}

Translation dynamics of the Brownian ellipsoids in the real space can be seen in the upper panel of Fig.~\ref{msd}, depicting mean squared displacements (MSD) for the isotropic, the nematic, and the liquid glass state. The bottom panel only shows MSDs for the liquid glass state $\Gamma=0.33$ and its parallel (MSD$_{\parallel}$) and perpendicular (MSD$_{\perp}$) components. The reference point for the latter two is the ellipsoid's orientation at t=0. The bottom panel shows MSD$_q$ obtained from the self intermediate scattering functions $F_s(q,t)$ using the Gaussian approximation explained in the figure caption. The upper panel of Fig.~\ref{msd} shows that the short time diffusion of all states is similar since the ellipsoids have not collided yet. Once the collisions start to take place around $t=2\times 10^{-4}$, the dynamics differs. At longer times arrest is seen as a weak plateau for the isotropic and nematic states ($\Gamma=0.16 - 0.25$), and a clear plateau in the liquid glass state $\Gamma=0.33$. At long times, particle motions again become diffusive for all states. The bottom panel of Fig.~\ref{msd} shows that in the liquid glass state $\Gamma=0.33$ the main contribution to the MSD plateau is due to the parallel motions of the ellipsoids. For small wave vectors, the self intermediate scattering function $F_s(q,t)$ can be used to obtain the mean squared displacement MSD$_q$. The MSD and MSD$_q$ curves show similar behaviour over all time scales which confirms that at short and long timescales in liquid glass, the particle translation motions are diffusive, while at intermediate times $10^{-1} < t < 10$ caging arrests the translation motion of the ellipsoids. 
\begin{figure}[h!]
     \centering
     \begin{subfigure}[b]{0.45\textwidth}
         \centering
         \includegraphics[width=\textwidth]
         {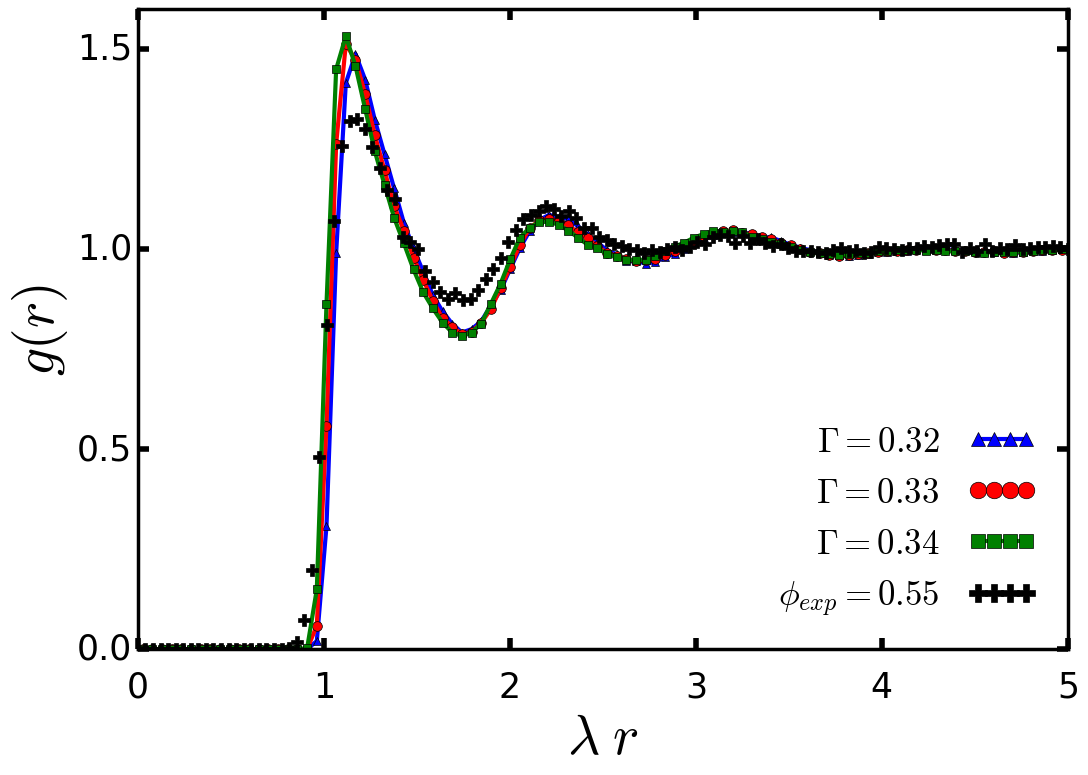}
         \caption{}
     \end{subfigure}
     \hfill
     \begin{subfigure}[b]{0.45\textwidth}
         \centering
         \includegraphics[width=\textwidth]
         {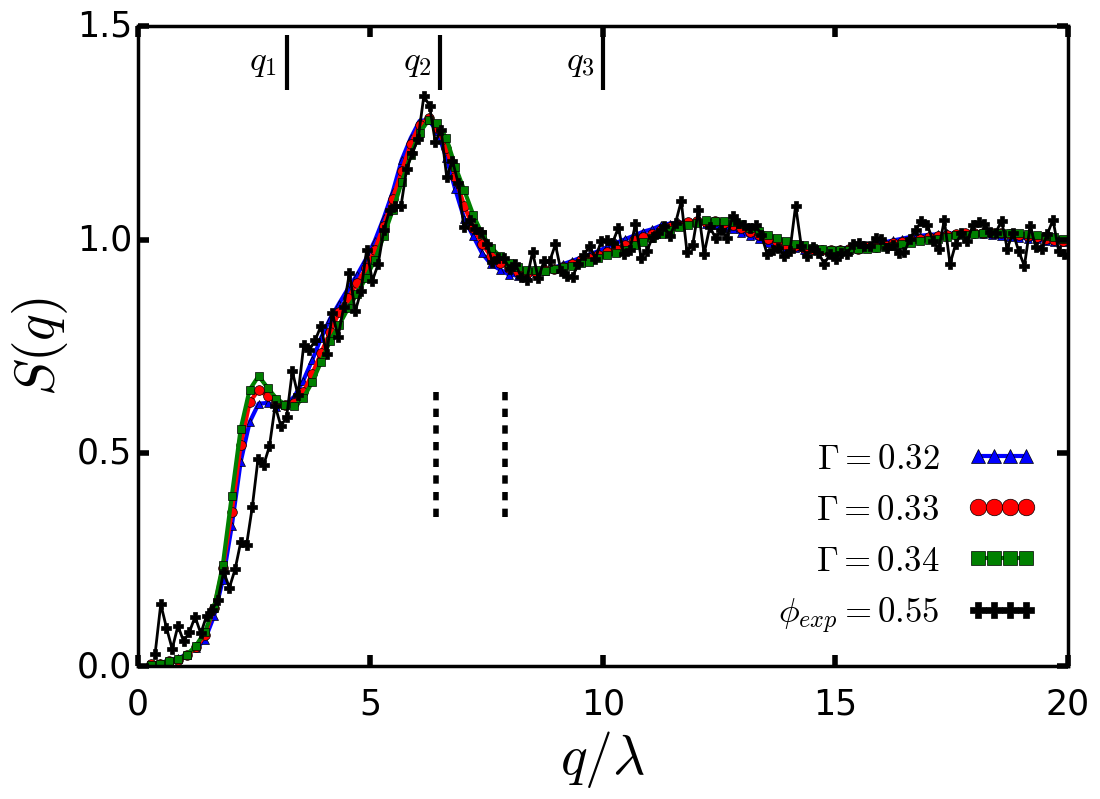}
         \caption{}
     \end{subfigure}
     \hfill
        \caption{The structural functions, $g(r)$ and $S(q)$, at different liquid glass densities. $\lambda$ is a scaling parameter which maps the simulation (soft-ellipsoid) curves to the experimental (hard-ellipsoid) curves. For the hard ellipsoids  $\lambda=1.0$ and for the soft ellipsoids $\lambda=1.04$. The  vertical dashed lines in $S(q)$ mark the wave vectors used in Fig.~\ref{sisf_ocf4_allGammas}, the vertical solid lines mark the $q$ whose dynamics is analysed according to MCT in Fig.~\ref{transCorr_fitting_33}.}
        \label{rdf_r_ordf_allGammas}
\end{figure}
\begin{figure}[h!]
     \centering
     \begin{subfigure}[b]{0.45\textwidth}
         \centering
         \includegraphics[width=\textwidth]
         {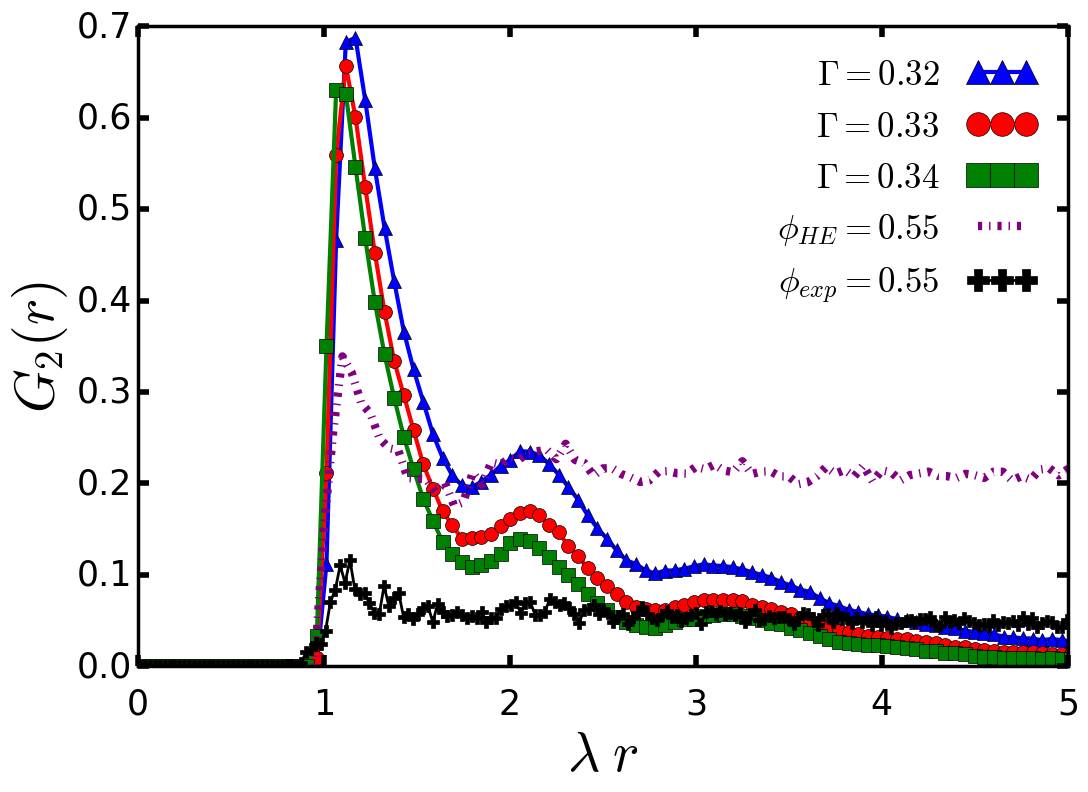}
         \caption{}
     \end{subfigure}
     \hfill
     \begin{subfigure}[b]{0.45\textwidth}
         \centering
         \includegraphics[width=\textwidth]
     {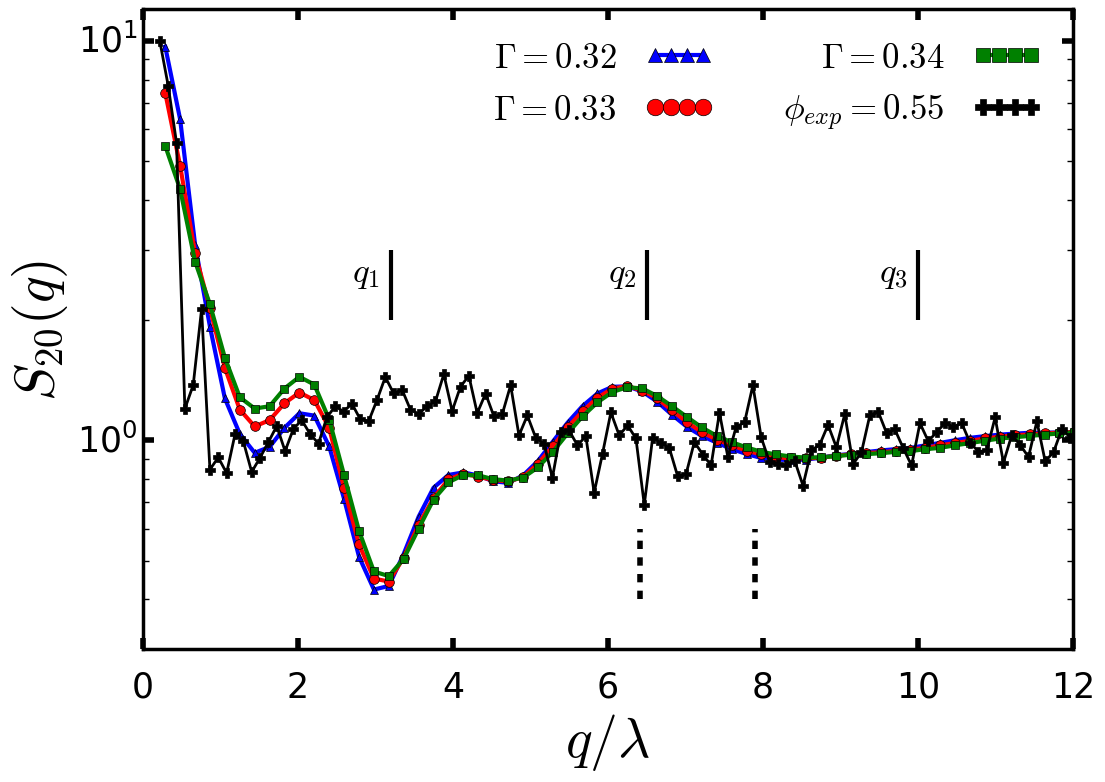}
         \caption{}
     \end{subfigure}
     \hfill
        \caption{The orientation structure functions, $G_2(r)$ and $S_{20}(q)$, at different liquid glass densities. $\lambda$ is a scaling parameter which maps the simulation (soft-ellipsoid) curves to the experiment (hard-ellipsoid) curves. The upper panel includes  the function $G_2(r)$ from a EDBD hard ellipsoid simulation. For the hard ellipsoid plots $\lambda=1.0$ and for the soft ellipsoid plot $\lambda=1.04$. In the lower panel, the vertical solid lines mark the $q$ whose dynamics is analysed according to MCT in Fig.~\ref{rotCorr_fitting_33}.}
       \label{rdf_ordf_allGammas}
\end{figure}

It is of interest to study the liquid glass at different densities. Figures \ref{rdf_r_ordf_allGammas} and \ref{rdf_ordf_allGammas} clarify the static structure functions of the liquid glass at different densities. In these figures the available experimental data are plotted. In the experiment, liquid glass was found at a volume fraction $\phi_{exp}=0.55$. As before, the factor $\lambda$ has the same meaning as in the isotropic phase in section \ref{ss:isotropicNematicSect}, but here its best value for mapping the soft-ellipsoid liquid glass to the hard-ellipsoid liquid glass is $\lambda=1.04$. From Fig.~\ref{rdf_r_ordf_allGammas} it is obvious that the three isotropic states whose densities in simulation are $\Gamma=0.32$, $\Gamma=0.33$ and $\Gamma=0.34$ map well onto the liquid glass state in experiment with volume fraction $\phi=0.55$. The $\lambda$ value suggests that the effective size of the soft and hard ellipsoids seems to be almost identical. The fact that both systems are at such high densities makes the packing constraints more important and similar, which results in $\lambda \approx 1$. This agreement can be checked from the plots of the functions $g(r)$ and $S(q)$ in Fig.~\ref{rdf_r_ordf_allGammas}, which agree except for the small peak at $q=2.7$ in the structure factor $S(q)$ in the simulation. This small peak indicates that there is a weak short-range translation order at this length scale in the soft-ellipsoid fluid. This weak translation order in the simulation cannot be observed in the $S(q)$ plot of the experiment (a weak indication of a shoulder can be noticed, however). It is directly related to the short range orientation order seen in the simulation plots of $S_{20}$ for the three densities; see Fig.~\ref{rdf_ordf_allGammas}. The two peaks in the simulation plot of $S_{20}(q)$ at $q/{\lambda} \le 4.5$ are a clear signal of the important role played by the ellipsoid softness discussed in the previous section. This short range orientation order is not seen in the experiment which reflects the fact that the orientation structure in the simulation does not exactly match the orientation structure in the experiment. A possible explanation for this is that the procedures of quenching the fluid into the liquid glass are not identical in simulation and experiment. In addition the ellipsoids in simulations interact via an anisotropic potential which clearly enhances the translation and orientation structures at short length scales. Another point to note in the simulation plot of $S_{20}$ is that its small wave vector limit decreases when the density increases. This is due to the fact that the isotropic-nematic transition density is further away at higher densities. Finally, the plot of $G_2(r)$ in Fig.~\ref{rdf_ordf_allGammas} shows that the liquid glass state does not show any kind of global nematic order and that even the local nematic order becomes weaker as the density increases.  This weakening of the global nematic order on increasing  the density can also be observed in the global order parameter $S$ shown in the lower panel of Fig.~\ref{diagQ_lg_m2m3m4_diffGammas}. Figure~\ref{rdf_ordf_allGammas} includes the $G_2(r)$ of hard ellipsoids from the experiment and from the EDBD simulations of small systems \cite{RollerLaganapan2020}; the latter show local alignment somewhat weaker but comparable to the soft ellipsoids, and otherwise are in a nematic state at the volume fraction where the dispersion forms a liquid glass. The $G_2(r)$ measured in the dispersion is very weak presumably for the reasons mentioned above when discussing $S_{20}(q)$.
\begin{figure}[h!]
     \centering
     \includegraphics[width=0.45\textwidth]{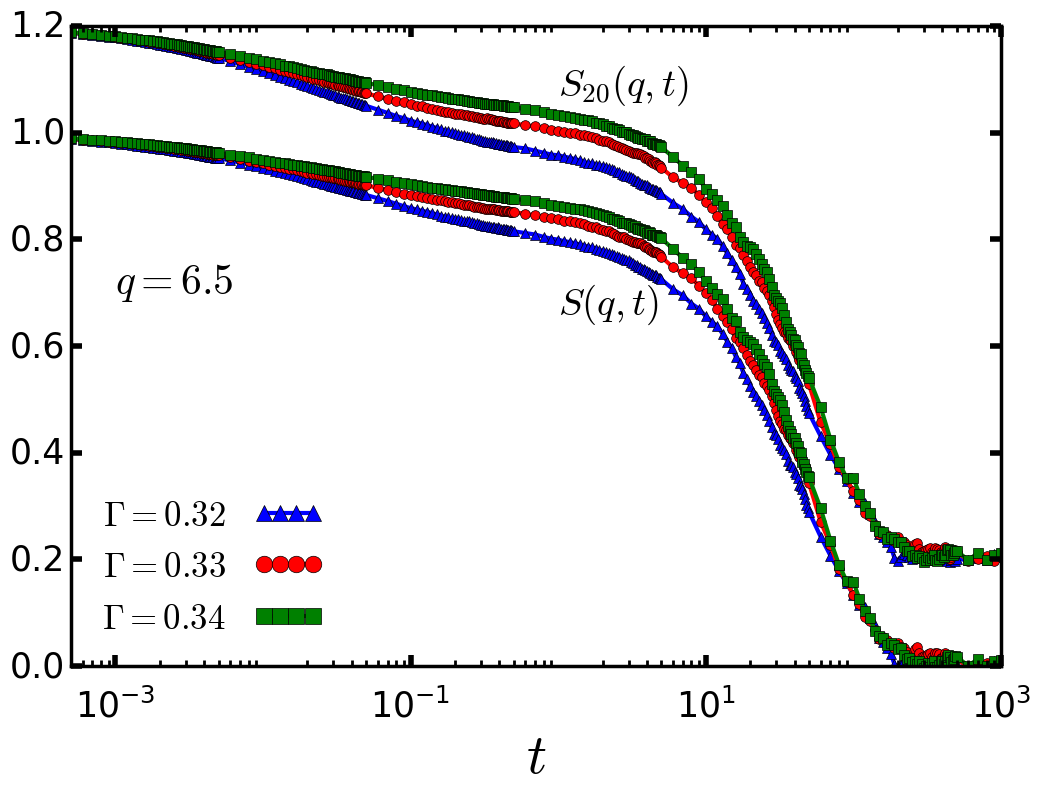}
     \hfill
        \caption{Dynamic structure and orientation structure factors, $S(q,t)$ and $S_{20}(q,t)$, from the simulation at different densities in the liquid glass state for $q=6.5$.}
       \label{dySq_dySq20_allGammas}
\end{figure}

\begin{figure}[h!]
     \centering
     \includegraphics[width=0.45\textwidth]{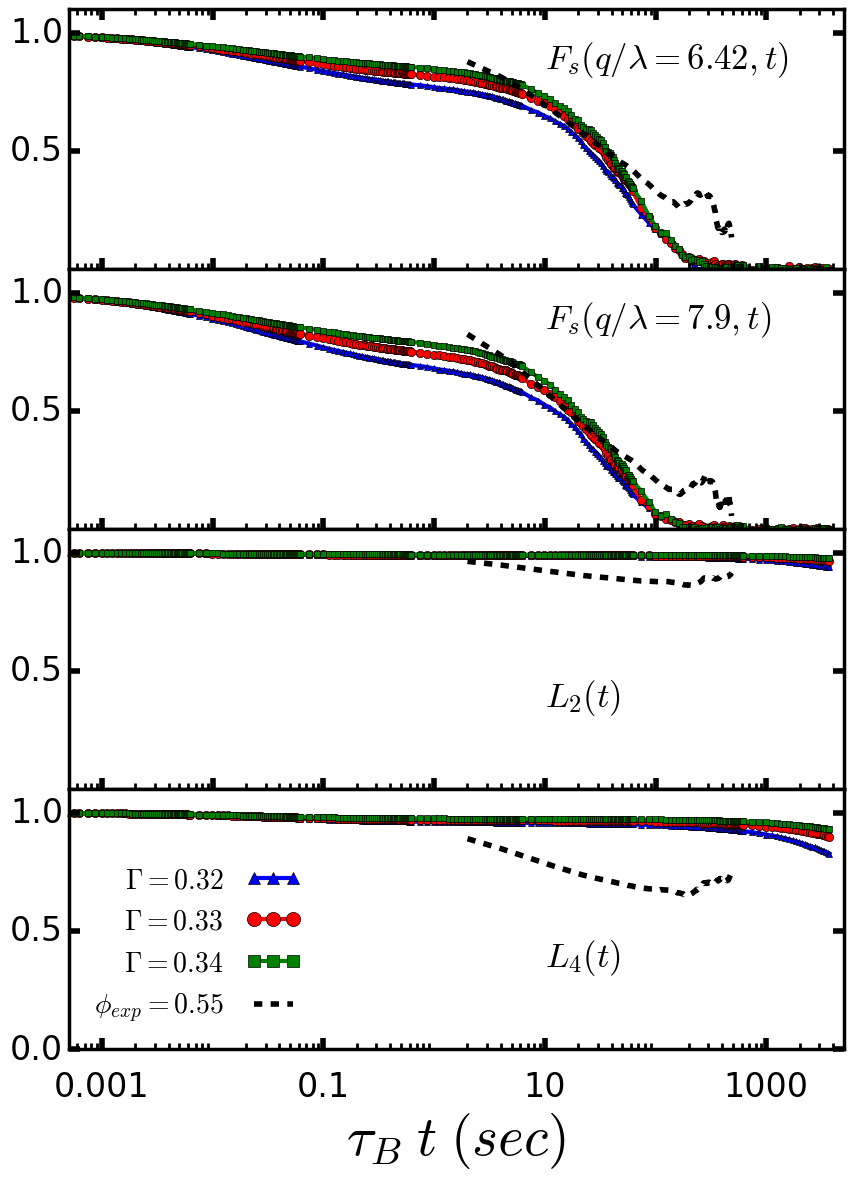}
     \hfill
        \caption{Self intermediate scattering function, $F_s(q,t)$ (for $q$ values marked in Fig.~\ref{rdf_r_ordf_allGammas}), and angular Legendre polynomial $L_2(t)$ and $L_{4}(t)$ at different densities in the simulation compared to the same functions in the experiment at volume fraction $\phi_{exp}=0.55$; parameters $\lambda$ taken from Figs.~\ref{rdf_r_ordf_allGammas} and \ref{rdf_ordf_allGammas}. The time scale is mapped via $\tau_B=1.24 \; sec$.}
       \label{sisf_ocf4_allGammas}
\end{figure}

The dependence of liquid glass dynamics on the density is illustrated in figures \ref{dySq_dySq20_allGammas} and \ref{sisf_ocf4_allGammas}. Figure \ref{dySq_dySq20_allGammas} only contains simulation curves of the functions $S(q,t)$ and $S_{20}(q,t)$ at $q=6.5$ which is the  \text{unscaled} location of the main peak in the simulation structure factor $S(q)$. The dependence of the translation dynamics on the density only appears during the lifetime of the cage for every dynamic function ($S(q,t)$ or $S_{20}(q,t)$). Unexpectedly, the translation correlations increase when the density increases only at timescales starting at the $\beta$-relaxation regime until $t=50$ in the $\alpha$-relaxation regime. After that, the density dependence disappears and the translation dynamics for the three states $\Gamma=0.32$, $\Gamma=0.33$, and $\Gamma=0.34$ behave identically, being characterized by a common timescale. This last finding does not agree with the MCT expectations since MCT predicts that the $\alpha$ relaxation timescale becomes longer when the density becomes higher \cite{Goetze,Letz2000,RollerLaganapan2020}. However, some MCT predictions can still be seen, e.g. the increase of the plateau height as the density increases. In this context it is relevant that MCT considers the relaxation of an equilibrium fluctuation while the simulation  records relaxations starting from a metastable state obtained after quenching.  

Figure \ref{sisf_ocf4_allGammas} shows the comparison between simulation and experiment in terms of the translation dynamics illustrated by $F_s(q/\lambda=6.42,t)$ and $F_s(q/\lambda=7.9,t)$ and the rotation dynamics illustrated by the functions $L_2(t)$ and $L_4(t)$. The experiment could only probe the long time dynamics of the correlation functions in this case. The wave vector $q/\lambda=6.42$ corresponds to the peak of the structure factor $S(q)$ in Fig.~\ref{rdf_ordf_allGammas} (the two wave vectors just mentioned are marked by the two vertical dashed lines in the $S(q)$ plot). The behavior of the dynamics of the function $F_s(q/\lambda,t)$ in the simulation is the same as the dynamical behavior of the functions $S(q,t)$ and $S_{20}(q,t)$ when it comes to the density dependence and the MCT predictions. Figure \ref{sisf_ocf4_allGammas} shows the best mapping in the dynamics when $F_s$ of the simulation is mapped at $\Gamma=0.33$ onto the experiment at $\phi_{exp}=0.55$. Both are liquid glass states. The Brownian timescale in this mapping is $\tau_B=1.24\;sec$ which is much smaller than the Brownian timescales in the isotropic phase. The reason why the liquid glass Brownian timescale is much smaller than the Brownian timescales in the previous section ($\tau^{iso}_B=140\;sec$ and $\tau^{iso}_B=220\; sec$ in fluid states at $\phi_{exp}=0.40$ and $\phi_{exp}=0.46$, respectively) could be that at high densities the effective particle distances are smaller and thus the relevant energy scale in the soft-potential is higher. Consequently, the corresponding collision rate would be higher and the particles would diffuse faster.

The close agreement observed in the translation dynamics is not seen in the orientation dynamics as the functions $L_2(t)$ and $L_4(t)$ illustrate. In the experiment, these two functions decorrelate to a lower value earlier than in the simulation. This is obvious from the decays of $L_2$ and $L_4$ in the experimental data around the timescale $t=50$ while the decays of $L_2$ and $L_4$ in the simulation occur at much later timescales which cannot be observed in the time window of the two panels in Fig.~\ref{sisf_ocf4_allGammas} anymore. The reason for the lower correlation at longer times in the experiment presumably is the inevitable noise in the experimental environment which was also seen in the isotropic dynamics in Fig.~\ref{sisf_ocf_iso}. Importantly, however, also the experimental dynamic functions show the absence of a final relaxation of angular correlations. At all of the densities and in the translation and orientation dynamics of the liquid glass, we observe that the rotation correlation functions persist at least two decades longer than the translation one.

\begin{figure}[h!]
     \centering
     \includegraphics[width=0.46\textwidth]{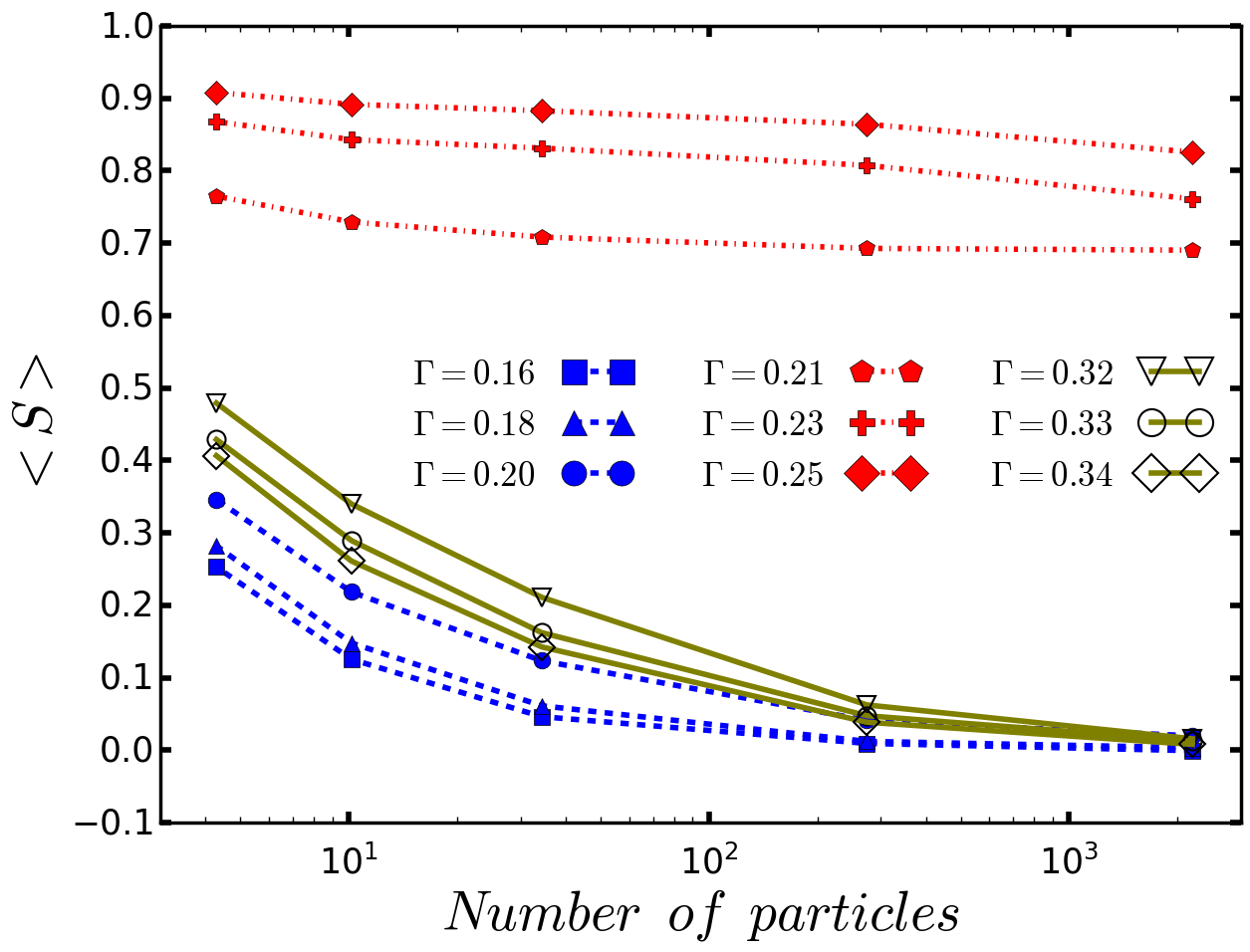}
     \hfill
        \caption{Cluster analysis of the local nematic order in boxes of a given number of particles for different states (see legend) of the simulated soft ellipsoid system. Nematic order quickly vanishes for bigger clusters in isotropic and liquid glass states; see text for details. The value of $S$ in the eight-box system (average number of particles $\approx 275$), is used in Figs.~\ref{ordfIsoNem} and \ref{sisf_ocf_iso_nem_lg}.}
      \label{clusterAnalysis}
\end{figure}

Finally, it is expedient to check how the nematic order changes as a function of the number of ellipsoids that form that nematic order. To this end, the simulation boxes are divided into a certain number of sub-boxes where each sub-box contains some number of ellipsoids. Then the nematic order parameter is computed in each sub-box and these values are averaged over the number of sub-boxes. The idea is to choose different numbers of sub-boxes leading to different numbers of particles per sub-box (or per cluster). Thereby, the average nematic order on different length scales  is computed. Figure \ref{clusterAnalysis} clarifies how the magnitude of the nematic order changes as the number of particles per box increases (or as the number of sub-boxes decreases) for several densities. The figure shows that for all densities there is some varying degree of nematic order when the cluster contains a small number of particles. Once the particle number per cluster starts to increase, the nematic order in isotropic and liquid glass states starts to disappear quickly while the nematic order in the liquid crystalline states stays significant. This clearly proves the absence of {\em compact} nematic domains in liquid glass. Rather, confocal microscopy found ramified nematic precursors \cite{RollerLaganapan2020}. For the nematic states (at $0.21\le \Gamma \le 0.25$), there is a drop in $S$ when it is calculated in  the total simulation box. This indicates some domain structure which would require even longer simulation runs to homogenize throughout the system; thus the value of $S$ for the eight-subbox system (average number of particles $\approx 275$) is used when comparing to $G_2(r)$ and $L_2(t)$ in Figs.~\ref{ordfIsoNem} and \ref{sisf_ocf_iso_nem_lg}, respectively. Results of the cluster analyses for the states $\Gamma=0.27,0.29,0.31$ are shown in the \hyperref[s:suppMater]{supplementary materials}. 

\subsection{MCT Analysis}
\label{s:MCT}
\begin{figure}[h!]
     \centering
     \begin{subfigure}[b]{0.46\textwidth}
         \centering
         \includegraphics[width=\textwidth]
         {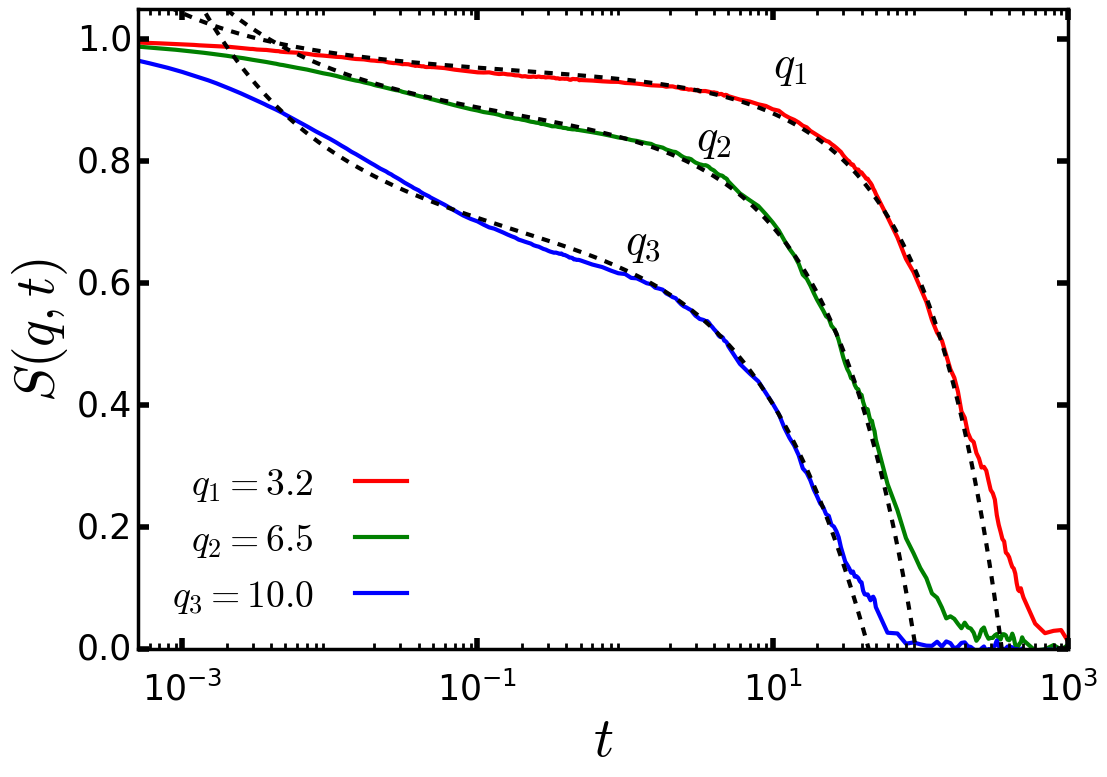}
         \caption{}
     \end{subfigure}
     \hfill
     \begin{subfigure}[b]{0.46\textwidth}
         \centering
         \includegraphics[width=\textwidth]
         {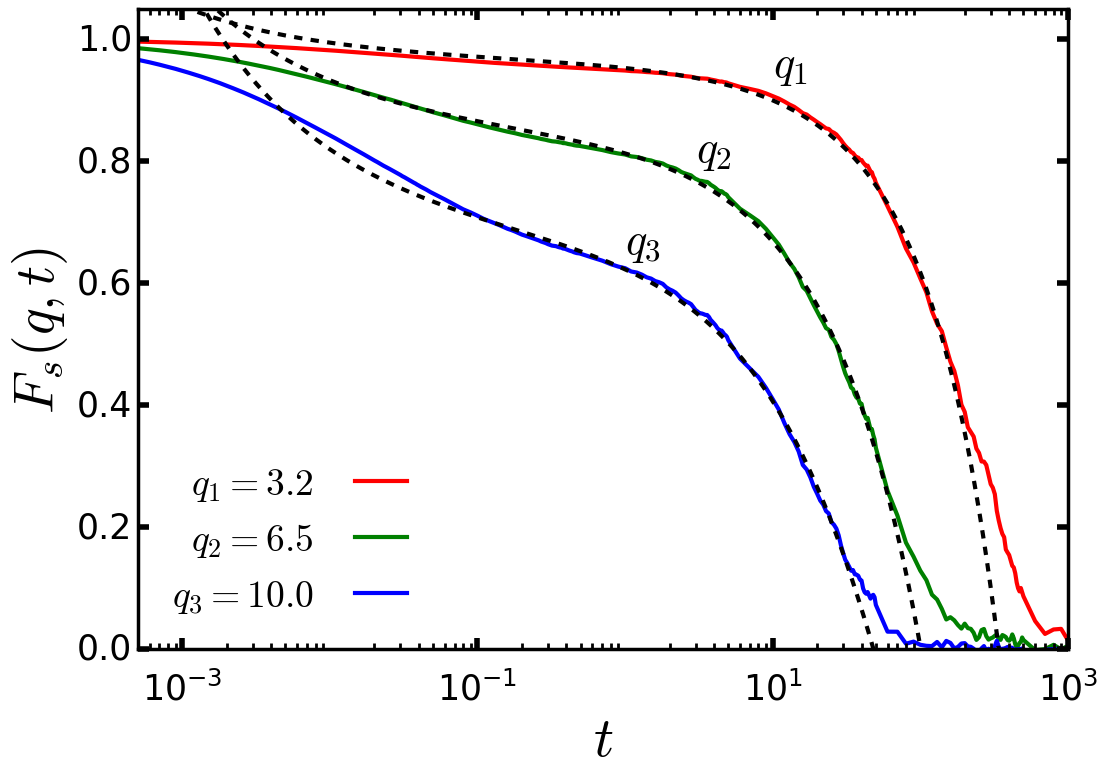}
         \caption{}
     \end{subfigure}
     \hfill
     \begin{subfigure}[b]{0.46\textwidth}
         \centering
         \includegraphics[width=\textwidth]
         {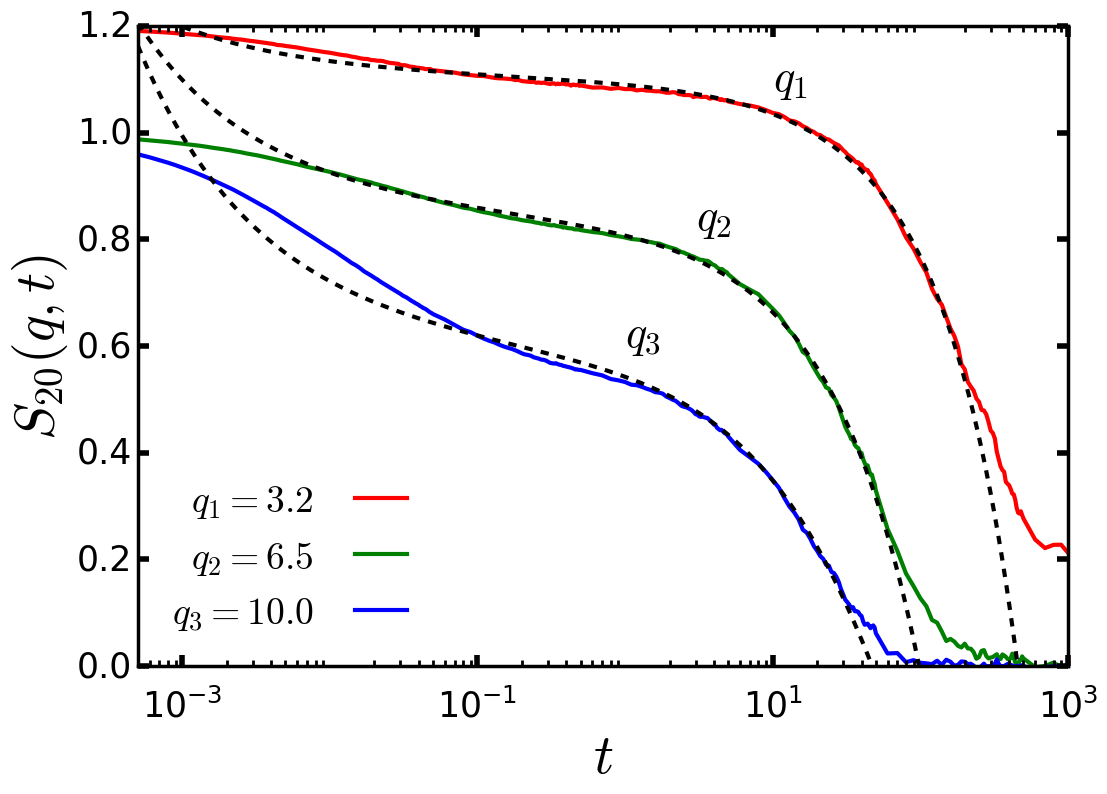}
         \caption{}
     \end{subfigure}
     \hfill
        \caption{The MCT fits to the translation correlation functions in the simulation at the density $\Gamma=0.33$ for the different wave vectors $q_1=3.2$, $q_2=6.5$ and $q_3=10$. $q_2=6.5$ is the location of the maximum peak in the structure factor $S(q)$. The $S_{20}(q,t)$ plot for the wave vector $q_1$ is shifted vertically for a better view. }
        \label{transCorr_fitting_33}
\end{figure}
Mode Coupling Theory (MCT) is the only microscopic theory to predict correctly many aspects of the structural relaxation close to the glass transition \cite{Goetze,Franosch-Fuchs,Letz2000}. It is based on solving the equations of motions for the correlation functions describing the dynamics of molecular fluids. The main input of the theory is the static structure factor of the fluid under study. Here, we mention the parts of the theory most relevant to our work. The MCT-analysis we perform, studies the incipient frozen-in correlations of the glass, how the system approaches it in the critical law (a power-law with exponent $-a$), and how it starts to decay during the von Schweidler law in the fluid (power-law with exponent $b$). In the glass, the long-time limit is a constant. The fitted MCT-laws are neither valid for short times, where colloids diffuse before colliding with others, nor for long times, where the liquid structure finally decorrelates following  a stretched exponential. According to Franosch \textit{et al.} \cite{Franosch-Fuchs}, the MCT predicts the time evolution of all correlation functions according to the following MCT solution
\begin{equation}\label{MCT-solution}
\begin{split}
 \Phi_l(t) = f^c_l + & h_l\sqrt{\sigma} \; g_{\pm}(t/t_{\sigma}) \\ 
 &+ h_l\sigma[h_{\pm}(t/t_{\sigma}) + K_l\;g_{\pm}(t/t_{\sigma})^2 \pm R_l]
\end{split}
\end{equation}
where the positive sign refers to a glassy state, e.g.~the rotation dynamics in liquid glass, while the negative sign refers to a fluid one, e.g.~the translation dynamics in liquid glass. The label $l$ becomes $q$ (which refers to a wavenumber) when considering the translation dynamics. It becomes the Legendre-polynomial degree $n$ when considering the rotation dynamics. The other quantities in Eq.~\eqref{MCT-solution} are the $\beta$-scaling function $g_{\pm}(t/t_{\sigma})$, which 
obeys \cite{Goetze,Franosch-Fuchs}
\begin{align}
 g_{\pm}(t/t_{\sigma} \to 0) &\to (t/t_{\sigma})^{-a} \\
g_{+}(t/t_{\sigma} \to \infty) &\to 1/\sqrt{1-\lambda} \\
g_{-}(t/t_{\sigma} \to \infty) &\to - B (t/t_{\sigma})^{b} 
\end{align}
for short and long rescaled times, respectively, and the correction functions
\begin{align}
 h_{+}(t/t_{\sigma}) &= \kappa(a) (t/t_{\sigma})^{-2a} - \tilde{\kappa}(a) \exp(-t/t_{\sigma}) \\
 h_{-}(t/t_{\sigma}) &= \kappa(a) (t/t_{\sigma})^{-2a} + \kappa(-b) B^2 (t/t_{\sigma})^{2b} 
\end{align}

\begin{figure}[h!]
     \centering
     \begin{subfigure}[b]{0.47\textwidth}
         \centering
         \includegraphics[width=\textwidth]
         {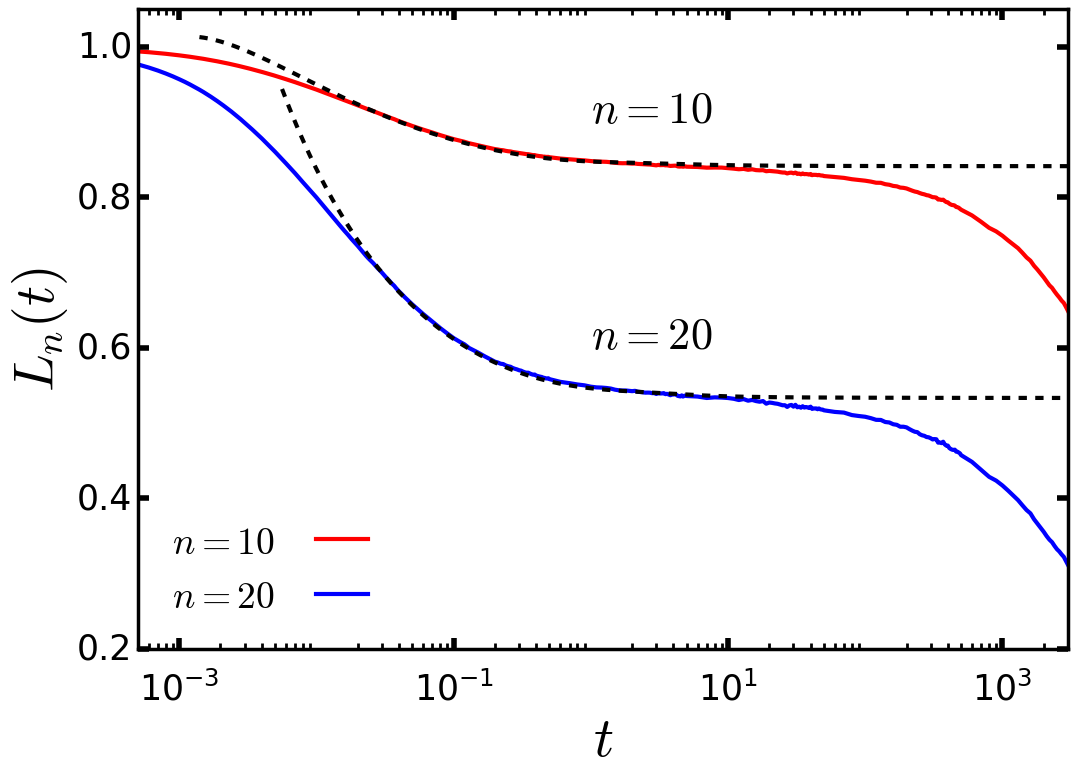}
         \caption{}
     \end{subfigure}
     \hfill
        \caption{The MCT fit to the orientation correlation functions in the simulation at the density $\Gamma=0.33$ for the order $n=10$ and $n=20$.}
        \label{rotCorr_fitting_33}
\end{figure}

These functions and parameters depend on the specific glass transition considered and hold for all correlators $\Phi_l$ \cite{Franosch-Fuchs}. The amplitudes $f_l, h_l, K_l$ and $R_l$ are $l$-dependent. $f^c_l$ gives an approximate estimate for the nonergodicity parameter at the singularity point (the glass transition density) while $h_l$ estimates the tightness of the cage arresting the dynamics. As we have two types of dynamics, we assert that there are two singularity points. The first is the rotation glass-transition density (or the fluid to liquid-glass transition density) labeled as $\Gamma^r_c$. The second singularity point is the translation glass-transition density labeled as $\Gamma^t_c$; it leads from liquid glass to (regular) glass \cite{RollerLaganapan2020}. According to MCT discussion in Ref.~\cite{Letz2000}, using the density as a control parameter, the liquid glass appears before the translation glass ($\Gamma^r_c < \Gamma^t_c$). We call all other parameters in the MCT solution, Eq. \eqref{MCT-solution}, the \textit{fitting parameters}. The separation fitting parameter is defined as $\sigma=\sigma_0 (\Gamma-\Gamma_c)/\Gamma_c$, where $\Gamma_c$ is a singularity point at which the MCT solutions bifurcate  ($\sigma=0$). $\sigma_0$ is some constant, and the scaling time is given by $t_{\sigma}=t_0/{\sigma}^{(1/2a)}$ where $t_0$ is a system-specific cross-over time. The remaining parameters depend on the correlation function, and their values will be mentioned below. We implement the MCT analysis at the liquid glass density $\Gamma=0.33$ since it exhibits a time-dependence close to the experiments. Figure \ref{transCorr_fitting_33} shows the results of the best fits of the translation dynamics to the translation part of Eq. \eqref{MCT-solution} at $\Gamma=0.33$. Figure \ref{rotCorr_fitting_33} shows the best fits of the orientation dynamics to the orientation part of Eq. \eqref{MCT-solution} at the same density. 

For analyzing the translation correlation functions, the fitting parameters are set equal to the ones for the supercooled hard-sphere fluid \cite{Franosch-Fuchs} except for the exponents ($a$ and $b$) and the parameter $\kappa(a)$, whose values are found to be different. Therefore, the translation-dynamics fitting parameters are  $\sigma=0.03$, $\kappa(a)=0.4$, $\kappa(-b)=0.569$, $B=0.836$, $t_0=0.020$ (leading to $t_{\sigma}=4.262$), $a=0.327$, and $b=0.641$ (corresponding to $\lambda_{\rm MCT}= 0.7$ according to the MCT exponent relation \cite{Goetze}). For the analysis of the rotation dynamics, the best fitting values are found to be $\sigma=0.066$, $\kappa(a)=0.961$, $t_0=0.007$, $t_{\sigma}=0.293$, and $a=0.364$.  For the rotation-dynamics analysis the value $\tilde{\kappa}(a)=2.48$ is still the values of the supercooled hard-sphere fluid. The amplitudes $f^c_l$ and $h_l$ in the MCT solution Eq. \eqref{MCT-solution} are discussed in details below. On the other hand, the amplitude $R_l$ takes small values and contributes to the MCT fits by improving the values of the correlation-function plateaus by $10\%$ at most.
\begin{figure}[h!]
\centering
\includegraphics[width=0.48\textwidth]{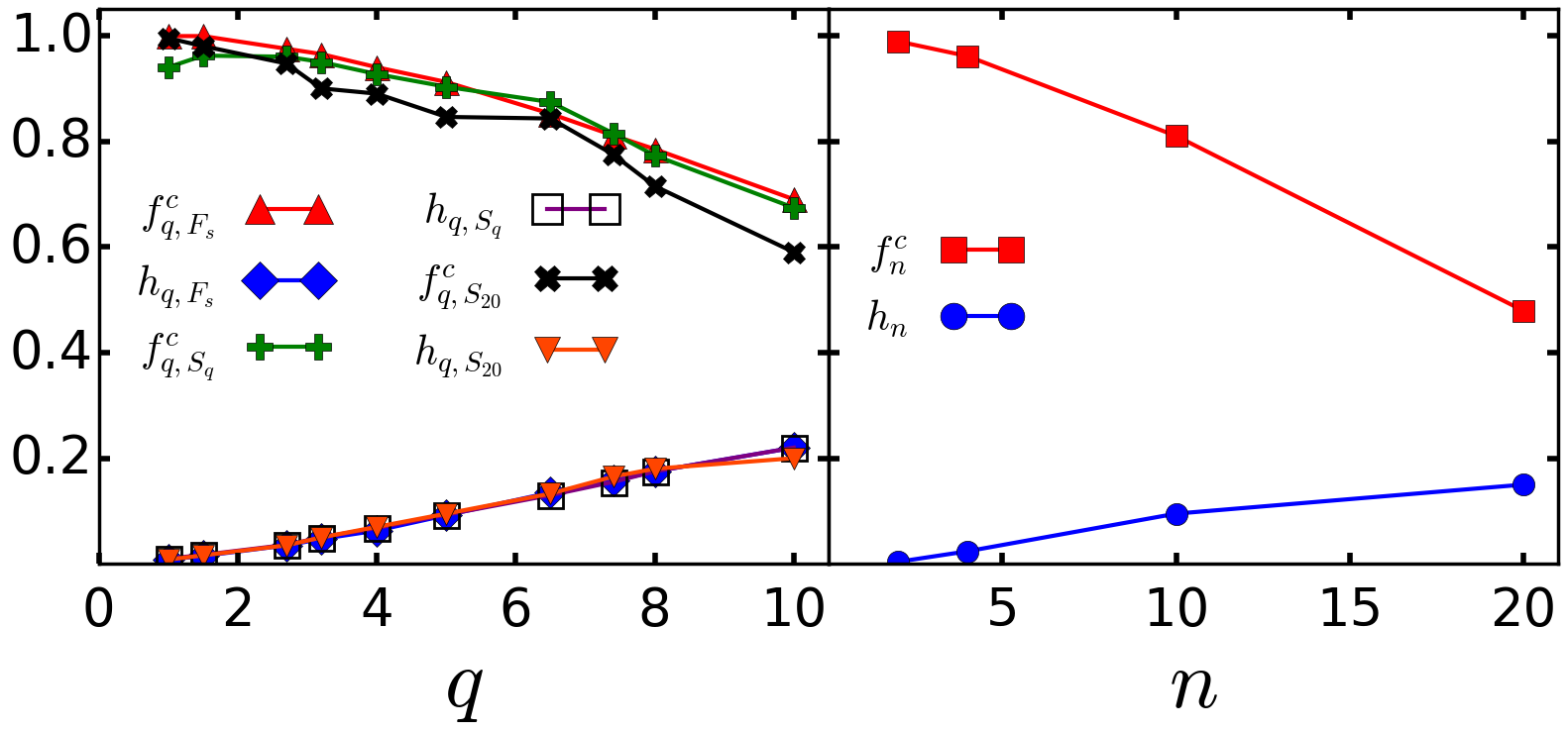}
\hfill
\caption{The MCT translation and rotation non-ergodicity parameters ($f^c_l$) and critical amplitudes ($h_l$) obtained from the MCT fits to the translation and rotation correlation functions at the density $\Gamma=0.33$ for different wave vectors and different Legendre-polynomial orders, respectively; for the fits see Figs.~\ref{transCorr_fitting_33} and  \ref{rotCorr_fitting_33}. The left panel gives the translation amplitudes and the right panel  the rotation amplitudes.}
\label{amplitudes_33}
\end{figure}

The exponents $a$ and $b$ are universal, meaning that they are the same at all length scales while fitting the translation correlation functions (Fig.~\ref{transCorr_fitting_33}). They are different for the orientation correlation functions but are constant at all Legendre orders $n$ while fitting the orientation correlation functions $L_n(t)$ (Fig \ref{rotCorr_fitting_33}). From Fig \ref{transCorr_fitting_33} we can notice that the translation correlation functions show agreement with MCT predictions around the turning point at the end of the plateau region  ($t\approx 1$) until some late timescales in the $\alpha$ relaxation regime, depending on the length scale we observe. Specifically, the MCT solution shows more agreement with the soft-ellipsoid translation dynamics at small wave vectors (note the curves at $q_1=3.2$ and $q_2=6.5$ in Fig.~\ref{transCorr_fitting_33}) than the translation dynamics at the large wave vectors (note the curve at $q_3=10.0$ in the same figure). The discrepancy between theory and simulations at the large wave vectors lies in the $\beta$ relaxation process. It is noteworthy that the long time decay in the translation dynamics is slower than predicted by MCT. It is expected that the discrepancies become smaller when the \textit{translation} glass transition is approached, i.e. when moving  closer to the translation bifurcation point of the MCT labeled as $\Gamma^t_c$. The values of the nonergodicity parameters $f^c_q$ and the critical amplitudes $h_q$ at different length scales are shown in the left panel of Fig.~\ref{amplitudes_33}. The nonergodicity parameter gives an \textit{estimate} of the frozen-in amplitude in the respective translation correlation function at the translation-glass singularity $\Gamma^t_c$. The decrease of $f^c_q$ values with the increase of the wave vectors indicates that the translation glass structure allows more motion at smaller length scale. The critical amplitudes $h_q$ for the three translation correlation functions are almost identical. This means that the tightness of the cage estimated by the three translation function at a certain wave vector is almost equivalent. The $h_q$ values increase as the wave vectors increase indicating that the glass structure of the ellipsoids gets arrested by a rather tight cage. Neglecting  quantitative differences and because most of our fitting parameters were taken from the ones of the supercooled hard-sphere fluid, we find that the translation dynamics of the soft-ellipsoid liquid glass state behaves similarly to the translation dynamics of the supercooled hard-sphere fluid.

The fits of the orientation dynamics depicted by $L_n(t)$ in Fig.~\ref{rotCorr_fitting_33} are discussed for two different orders $n=10$ and $n=20$. The higher the order, the weaker the orientation correlation becomes with time (even in liquid glass). It is clear from the figure that the MCT solution agrees with the dynamics in the $\beta$-relaxation regime. The solutions do not show any agreement with the curves in the region where the aging-relaxation starts ($t \approx 100$). The fact that at a comparable time for all $L_n(t)$ no agreement between the MCT solution and the $L_n$ curves can be found any more, is a signal of the break-down of the theory. The simulations clearly showed the aging of the orientation dynamics; see Fig.~\ref{orientCorr_m2m3m4_33}. Here, we analytically assert that the cooperative rotations within their orientational cage in the liquid glass state at $\Gamma=0.33$ move according to the MCT intermediate-time dynamics of glass. The nonergodicity parameter $f^c_n$ and the critical amplitude $h_n$ are shown in the right panel of Fig. $\ref{amplitudes_33}$. The value of $f^c_n$ gives an approximate value for the parameter at the rotation singularity point $\Gamma^r_c$. The $f^c_n$ values become smaller when the order $n$ increases. This tells that the orientation correlation becomes weaker when the rotation correlations are probed more locally. The critical amplitude $h_n$ illustrates how much the cage affects the orientation dynamics. At higher order $n$, the orientation cage of the ellipsoids exhibits larger motional amplitudes.

\section{Conclusions}
\label{s:Conclusion}

We performed large scale Brownian dynamics simulations of ellipsoidal particles (at aspect ratio 3.5) including dense systems quenched into {a partially arrested isotropic state. It is the first realization in simulations of} the recently discovered liquid glass state. Based on our time- and wave vector-windows that appreciably extend the experimentally accessible ones, we tested the predicted decoupling of translation and orientation motion. Glassy arrest of the angular correlations was evidenced by observing aging, while translation correlations relaxed in (supercooled) equilibrium and at least a factor of one hundred faster.  Our simulation results compare well to the experimentally measured structure and to the translation motion in the colloidal dispersions where liquid glass was first observed \cite{RollerLaganapan2020}. While the long-wavelength static orientational correlations agree between simulation and experiment, the local alignment of the elongated particles showed differences. Presumably as a consequence, the time-dependent orientational correlations agreed in their final relaxation or when arresting, but showed differences for shorter times. In the simulations, angular correlations remained higher for longer, which might be caused by the fact that the pairwise interaction in simulations is anisotropic.

By a detailed analysis of the power-law relaxation in the intermediate time dynamics, we tested whether mode coupling theory correctly describes the diffusion and orientation dynamics of the liquid glass state. Overall we found good agreement of the time-dependent structural functions over typically four orders in time-variation. Note, that we ignored two failures of the microscopic MCT. First, it is  well-known that the MCT glass transitions are rounded \cite{Goetze}. We identified the liquid glass state by aging thereby bypassing the question whether the final relaxation time is actually infinite or just longer than the simulation time.  The second aspect concerns the more specific prediction in Ref.~\onlinecite{Letz2000} that translation motion in liquid glass exhibits a small non-ergodic amplitude, while we find it to vanish. We follow the interpretation of the authors and previous tests in two \cite{Zheng2014,Zheng2011, Mishra2014,Sokolowsky2016,Wang2022} and three \cite{RollerLaganapan2020} dimensions to identify liquid glass with the state where rotation motion freezes while the structure on the average particle separation remains fluid. In the orientational dynamics we could only test the decay onto an intermediate time plateau because the final aging-induced relaxation is beyond the theory. Quenching into metastable states did not allow us to explore the change of the dynamics inside the liquid glass states as predicted by theory. Quenching to final glass states at higher effective density did not slow down translation motion. Presumably, the system falls  out off-equilibrium during the quenching process in a way not affected by the final density aimed for.

The strong orientational fluctuations that we recorded in the static angular structure factor support the prediction by Letz, Latz and Schilling \cite{Letz2000}, that liquid glass originates in nematic fluctuations, which become the entities of the mutual hindrance and of glassy arrest. During liquid glass formation, the local alignment of ellipsoids gets frustrated on intermediate length scales and gives ramified structures, viz.~differently oriented clusters. The emerging nematic precursors contain more and more particles, and it is the mutual obstruction of such cooperative regions that leads to the formation of liquid glass. For future work it would be interesting to study larger simulation systems and and include hydrodynamic interactions in order to slow down the formation of global nematic order more so that the transition to the liquid glass state, where diffusion proceeds but angular motion freezes, can be studied using computer simulations.

\section{Supplementary Materials}
\label{s:suppMater}
Refer to the supplementary materials which show the comparisons of the structure and dynamics in the liquid glass and the nematic states preceding the nematic-to-liquid-glass transition.

\begin{acknowledgements}
We thank Aleena Laganapan for discussions and help in the initial stages of the project.  The work was supported  by the Deutsche Forschungsgemeinschaft (DFG) via SFB 1432 project C07. 
\end{acknowledgements}

\bibliography{PaperBib}
\newpage \clearpage

\pagestyle{empty}

\begin{center}
\textbf{\Large{Supplementary Materials}}
\end{center}
\noindent \textbf{Observation of Liquid Glass in Molecular Dynamics Simulations}\\
Mohammed Alhissi, Andreas Zumbusch, Matthias Fuchs

\subsection*{Results for the densities $\Gamma=0.27, 0.29, 0.31$}

In the following, we report results for the states with $\Gamma=0.27, 0.29, 0.31$.  At $\Gamma=0.27$ and $\Gamma=0.29$, the system is in a nematic state, whereas the state at $\Gamma=0.31$ is a liquid glass state. Structure and dynamical functions are those defined in section III of the main article. The same simulation setup as described in section IV of the main article was used, except for the state with $\Gamma=0.29$. The latter required longer equilibration and measurement times than the other nematic states. Its equilibration ran until the scalar nematic order parameter fluctuated around its average value. The measurement time of the state $\Gamma=0.29$ was $10\times10^6\;dt^*$ with $dt^*=0.0001$. Since none of the plots in these supplementary materials contains experimental data, the mapping parameter $\lambda$ does not appear in the figures below.
\begin{figure}[h!]
     \centering
     \begin{subfigure}[b]{0.45\textwidth}
         \centering
         \includegraphics[width=\textwidth]
         {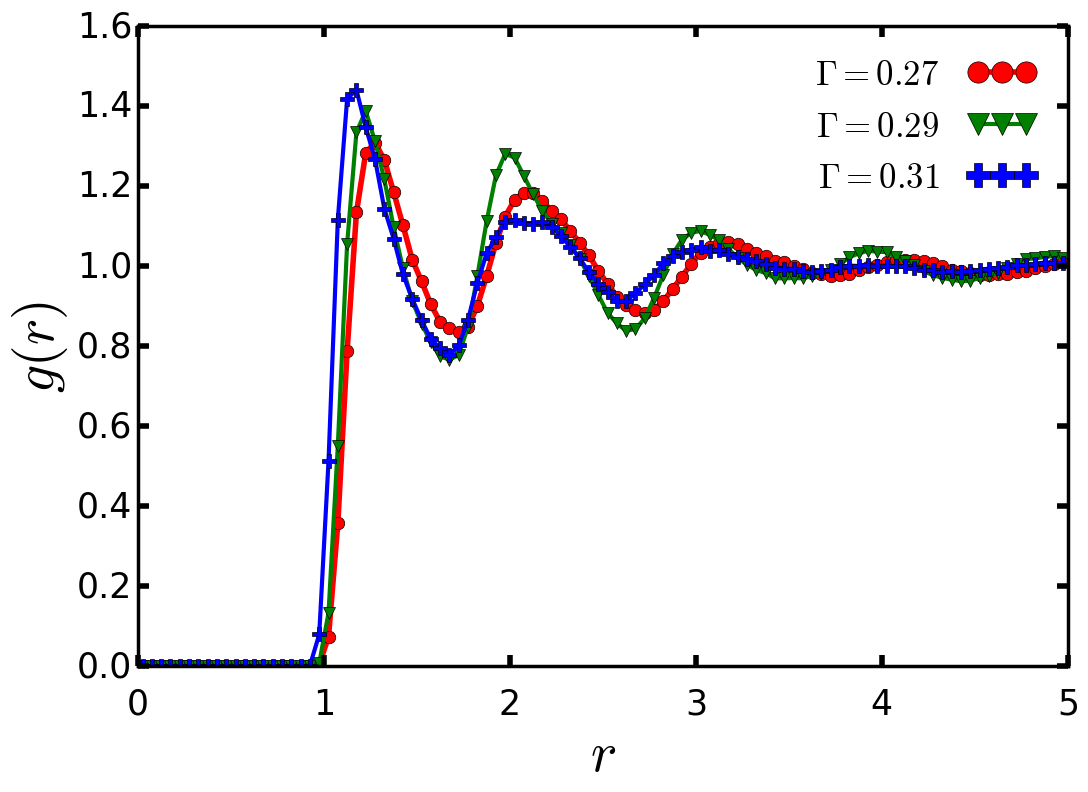}
     \end{subfigure}
     \begin{subfigure}[b]{0.45\textwidth}
         \centering
         \includegraphics[width=\textwidth]
         {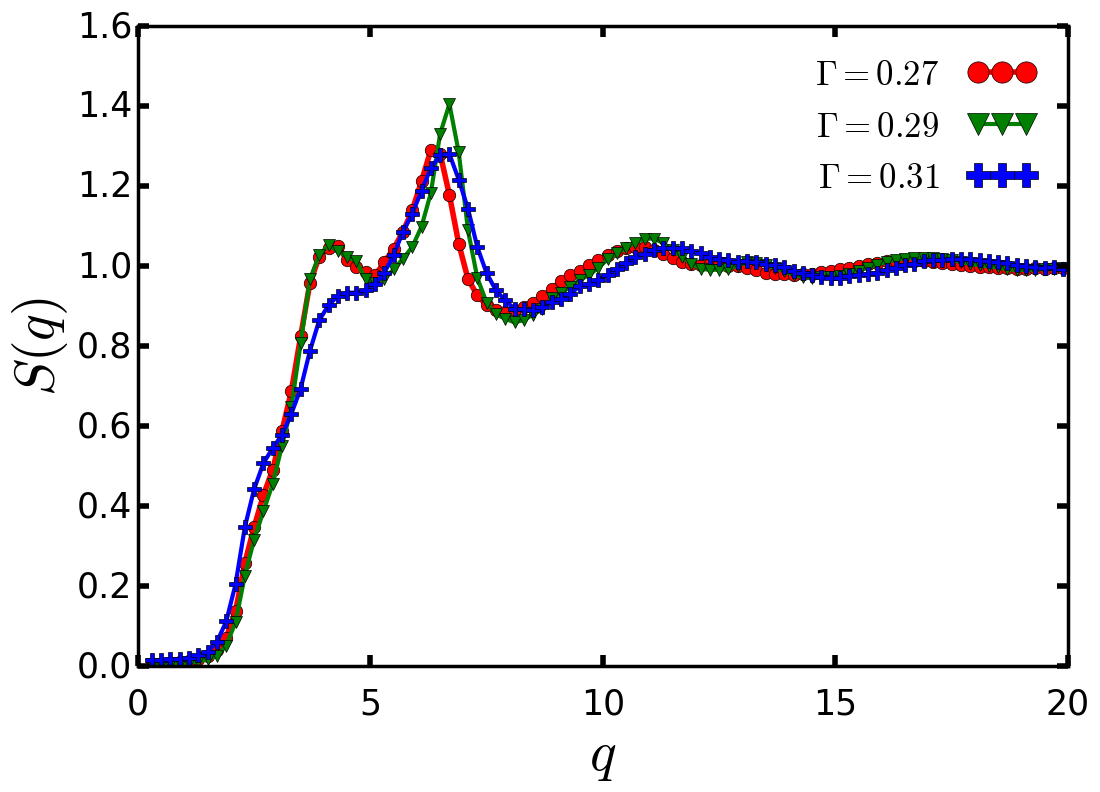}
     \end{subfigure}
\caption{The pair correlation function $g(r)$ and the structure factor $S(q)$ for the soft ellipsoid system for the nematic states $\Gamma=0.27$ and $\Gamma=0.29$ and for the liquid glass state $\Gamma=0.31$. No hard ellipsoid (experiment) data are shown here.}
\label{rdfSq_272931}
\end{figure}

The pair correlation $g(r)$ and the structure factor $S(q)$ functions giving insight into the translation order in the soft ellipsoid fluid for the states $\Gamma=0.27, 0.29 ,0.31$ are shown in Fig.~\ref{rdfSq_272931}. As for the densities described in the main article, increasing the effective density leads to an increase in the ellipsoids' local translation order. This can be seen in the first peaks of the pair distribution function $g(r)$ (Fig.~\ref{rdfSq_272931}). However, one also notices that the long range order in the plot of $g(r)$ for the  state $\Gamma=0.31$ is weaker than for the nematic states $\Gamma=0.27, 0.29$. This is expected as the ellipsoid alignment in the nematic states enhances the translation order while in the liquid glass there is no long range alignment. Plotting the structure factor $S(q)$ (Fig.~\ref{rdfSq_272931}), we find that the small bump at $q < 5$ is lower in the liquid glass than in the nematic states. Also, the amplitude of the most intense peak in the $S(q)$ plot is different from density to density. For the nematic phase, the first peak increases with the density but for the liquid glass the first peak is smaller than the first peak in the other nematic states. The fact that the nematic alignments do not extend for long distances in the liquid glass (see Fig.~\ref{clusterAnalyses2729}) makes the translation order smaller than the one in the nematic states at small wave vectors.

\begin{figure}[h!]
     \centering
     \begin{subfigure}[b]{0.45\textwidth}
         \centering
         \includegraphics[width=\textwidth]
         {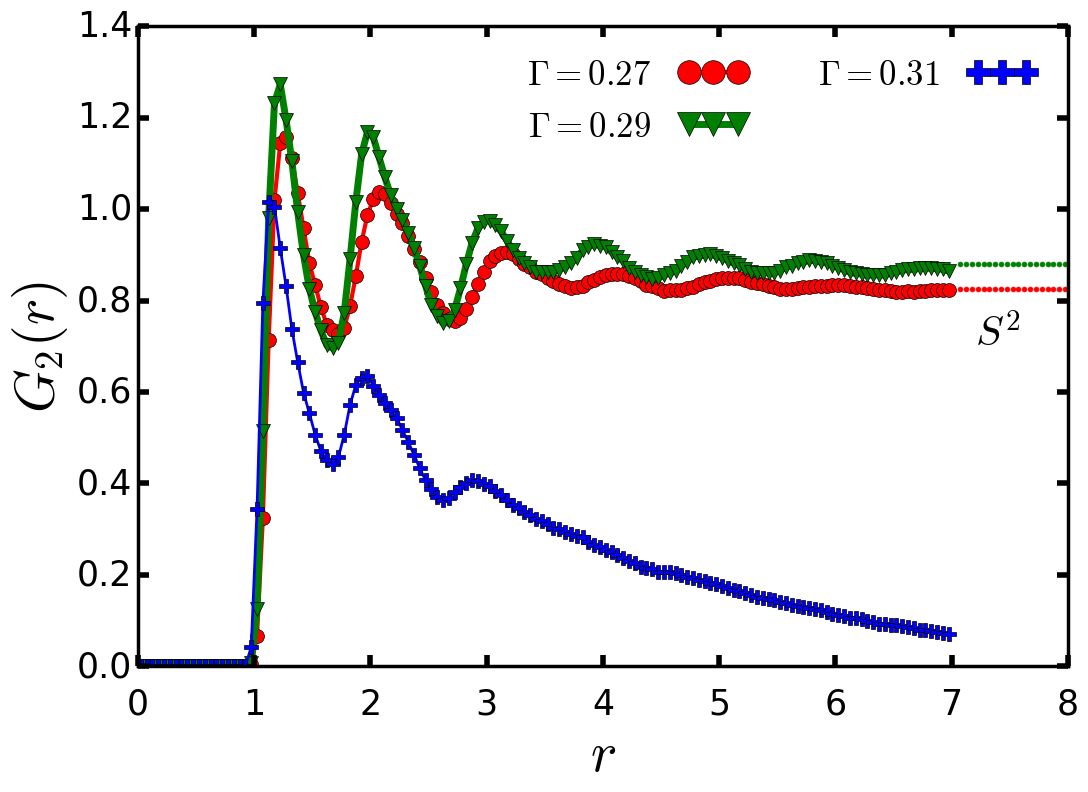}
     \end{subfigure}
     \begin{subfigure}[b]{0.45\textwidth}
         \centering
         \includegraphics[width=\textwidth]
         {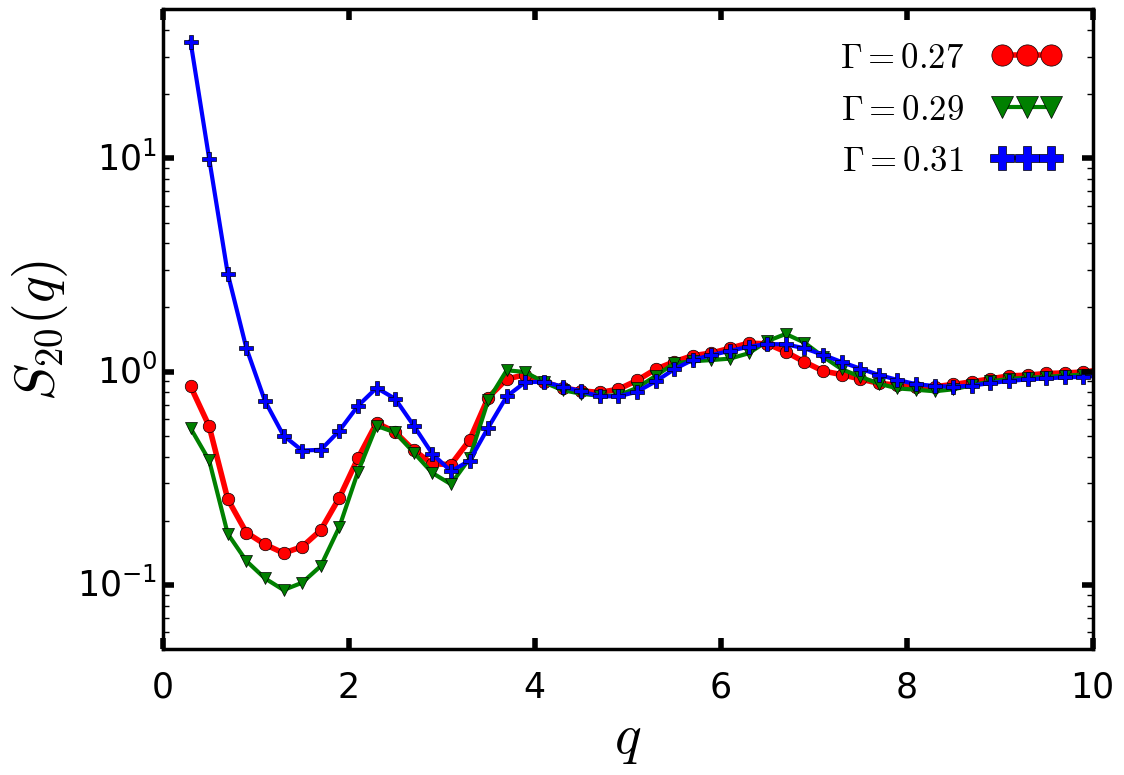}
     \end{subfigure}
\caption{The orientation pair correlation function $G_2(r)$ and the orientation structure factor $S_{20}(q)$ for the soft ellipsoid system for the nematic states $\Gamma=0.27$ and $\Gamma=0.29$ and for the liquid glass state $\Gamma=0.31$. $S^2$ shown in the plot of $G_2(r)$ is the quadratic value of the scalar nematic order parameter when the cluster size is $34$ particles (see Fig.~\ref{clusterAnalyses2729}).}
\label{oRDFSq20_272931}
\end{figure}
Functions giving insight in the orientation structures for the three states are shown in Fig.~\ref{oRDFSq20_272931}. The orientation pair distribution function $G_2(r)$ indicates the nematic order for the states $\Gamma=0.27,0.29,0.31$. In the liquid-glass state $\Gamma=0.31$, the drop of the function as the center to center distance $r$ increases is apparent. For the two nematic states $\Gamma=0.27,0.29$ the nematic order persists and the peaks in the plots of these two states show strong nematic order. For large $r$, $G_2(r)$ corresponds to the quadratic value of the scalar nematic order parameter $S$. Here, we take the nematic order parameter that corresponds to a cluster of size 34 ellipsoids (see Fig.~\ref{clusterAnalyses2729}) since equilibrating the whole nematic states was too time consuming. For the nematic states, we notice that the peaks of $G_2(r)$ for the state $\Gamma=0.29$ are higher than the peaks for the state $\Gamma=0.27$ at all length scales. This indicates that the nematic order becomes more enhanced as the nematic density increases. The second orientation structure function is the orientation structure factor $S_{20}(q)$ given in Fig.~\ref{oRDFSq20_272931}. Differences between the three states in the $S_{20}(q)$ plot appear at small wave vector values around $q<4$. One notices that when increasing the density, the values of $S_{20}(q)$ at very small wave vectors become larger. At wave vectors $q<1$, the $S_{20}(q)$ values are highest for the liquid glass state at $\Gamma=0.31$. This indicates that the fluctuations of the local orientation microscopic densities extend to the whole simulation box and a phase transition takes place in which the global nematic order is destroyed. 

\begin{figure}[h!]
\centering
\includegraphics[width=0.45\textwidth] 
{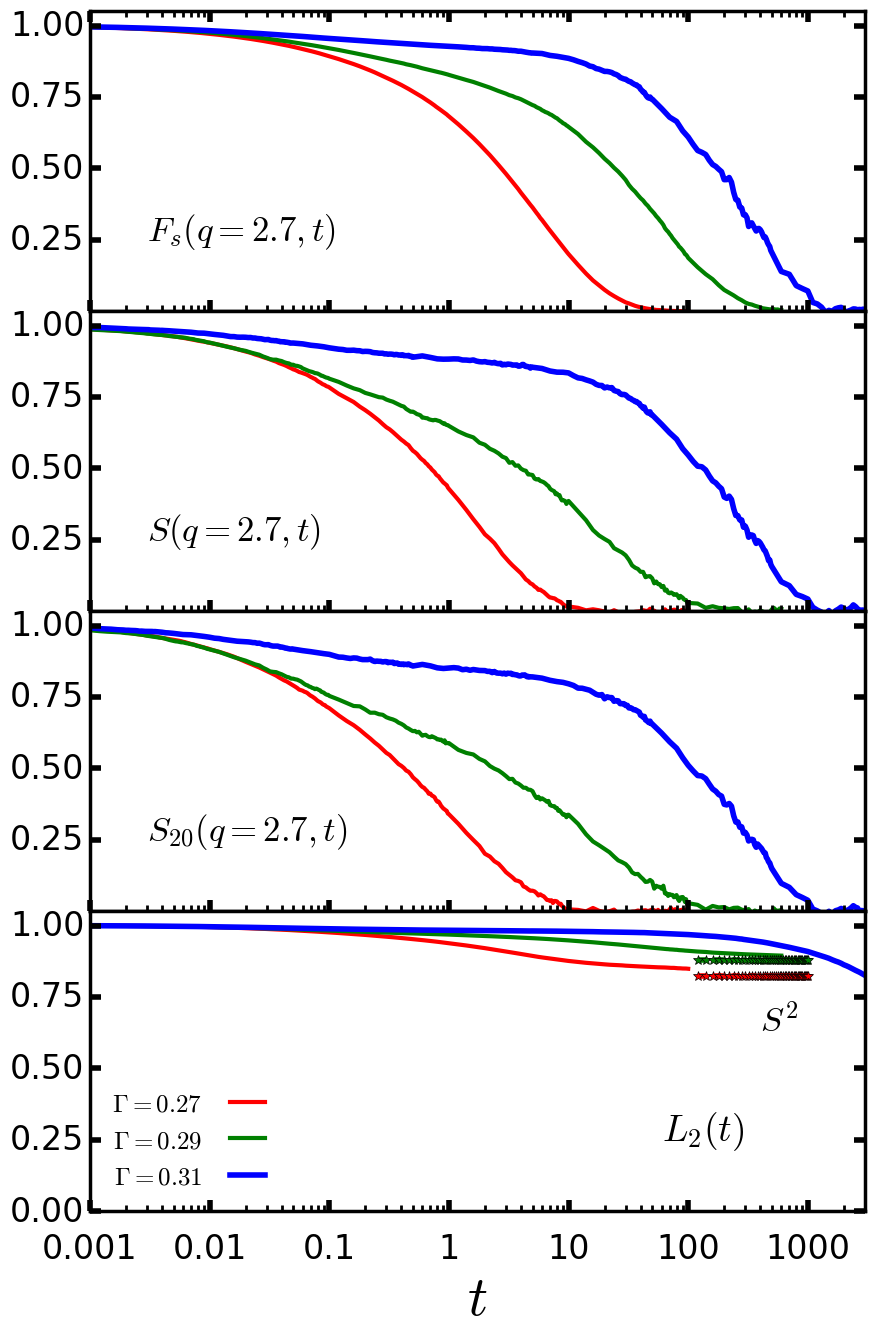}
\caption{The self intermediate scattering function $F_s(q=2.7,t)$, the dynamic structure factor $S(q=2.7,t)$, the orientation dynamics structure factor $S_{20}(q=2.7,t)$, and the second order orientation correlation functions $L_2(t)$ for the soft ellipsoid system for the nematic states $\Gamma=0.27$ and $\Gamma=0.29$ and for the liquid glass state $\Gamma=0.31$. $S^2$ shown in the plot of $L_2(t)$ is the quadratic value of the scalar nematic order parameter $S$ when the cluster size is $34$ particles (see Fig.~\ref{clusterAnalyses2729}).}
\label{trans&rot_272931}
\end{figure}

The dynamics of the three states $\Gamma=0.27,0.29,0.31$ can be inferred from the translation correlation functions $F_s(q=2.7,t)$, $S(q=2.7,t)$, $S_{20}(q=2.7,t)$ and from the orientation correlation function $L_2(t)$ in Fig.~\ref{trans&rot_272931}. The wave vector $q=2.7$ is close to the first peak in the plots of $S_{20}(q)$ in Fig.~\ref{oRDFSq20_272931}. The three translation correlation functions relax to equilibrium. They decay at longer times when the density increases. For the orientation correlation function $L_2(t)$, the nematic states decay to the quadratic values of their scalar nematic order parameters $S$ corresponding to the values of $S$ computed for a cluster size of 34 ellipsoids shown in Fig.~\ref{clusterAnalyses2729}. The reason for not choosing the cluster with the largest number of ellipsoids is that equilibrating the whole nematic samples at such high densities takes a lot of time. On the other hand, the function $L_2(t)$ for the state $\Gamma=0.31$ does not show any relaxation to equilibrium even though its scalar nematic order parameter $S$ vanishes for the cluster with the largest number of ellipsoids as shown in Fig.~\ref{clusterAnalyses2729}. In the nematic states, for all cluster sizes the scalar nematic order parameters $S$ are significant. Therefore, we identify $\Gamma=0.31$ as the first liquid glass state in our Brownian dynamics simulations of soft ellipsoid fluid with aspect ratio $\eta=3.5$. We conclude that the liquid glass transition occurred in the density range $0.29<\Gamma<0.31$.

\begin{figure}[h!]
\centering
\includegraphics[width=0.4\textwidth] 
{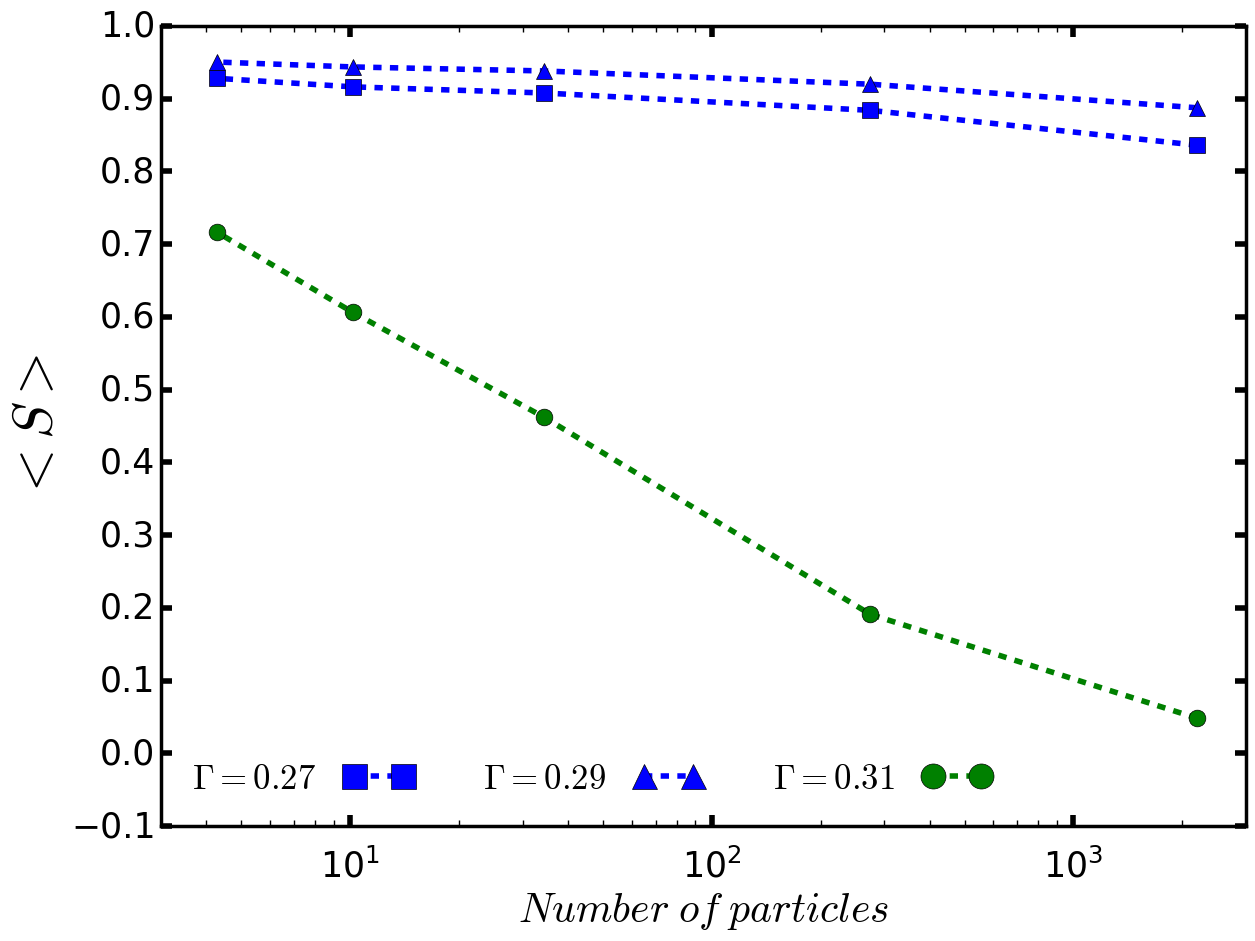}
\caption{Cluster size analysis for the nematic states (blue) $\Gamma=0.27$ and $\Gamma=0.29$ and for the liquid glass state (green) $\Gamma=0.31$.}
\label{clusterAnalyses2729}
\end{figure}
\begin{figure}[h!]
\centering
\includegraphics[width=0.4\textwidth] 
{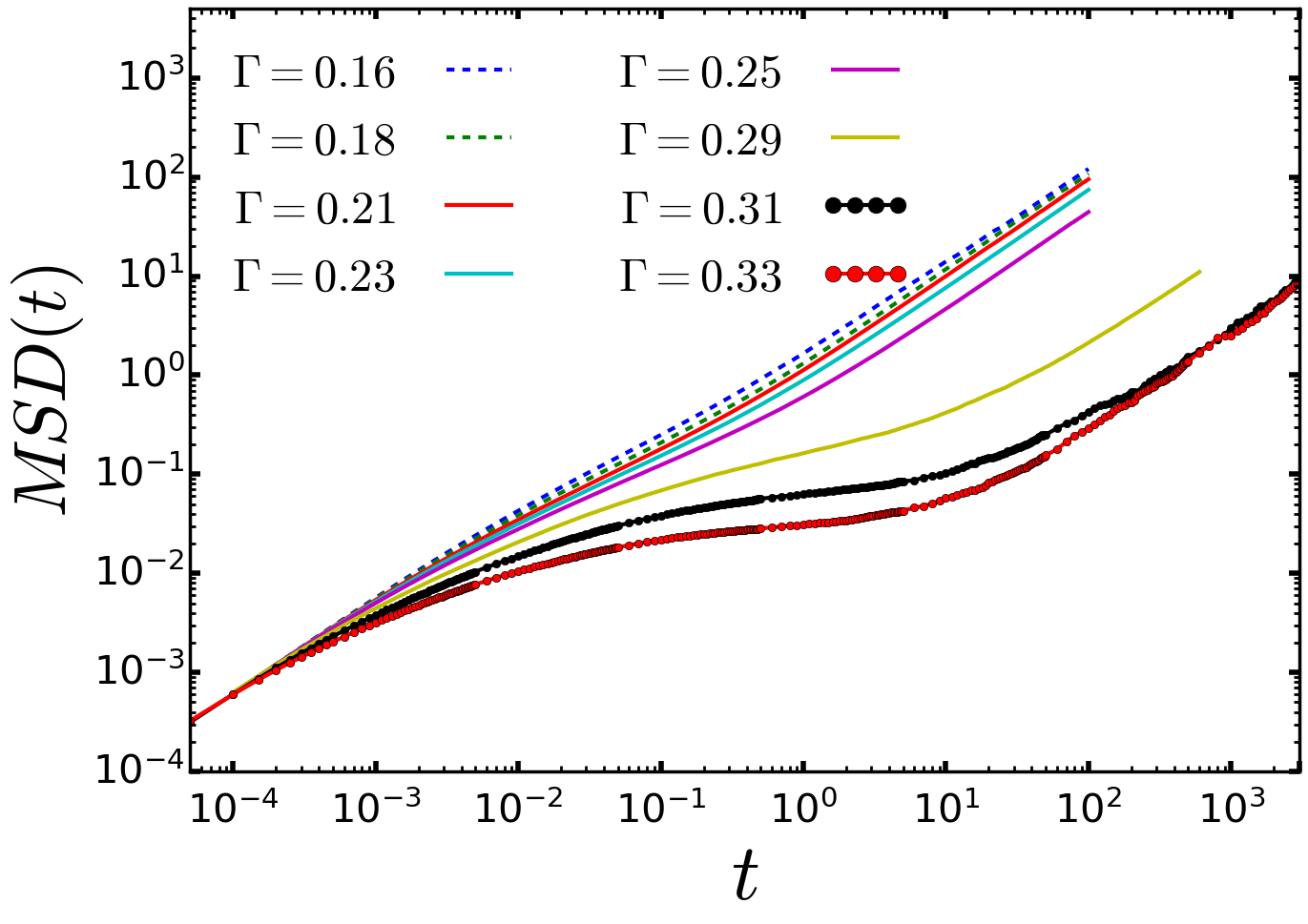}
\hfill
\caption{Mean squared displacements (MSD) for isotropic ($\Gamma=0.16, 0.18$), nematic ($\Gamma=0.21 - 0.29$), and liquid glass ($\Gamma=0.31, 0.33$) states.}
\label{msd2}
\end{figure}
Fig.~\ref{msd2} shows the mean squared displacements (MSD) for the isotropic states $\Gamma=0.16, 0.18$, the nematic states $\Gamma=0.21 - 0.29$, and the liquid glass states ($\Gamma=0.31, 0.33$). In all states, the Brownian ellipsoids diffuse identically at very short time scales before collisions start to take place at $t \approx 2 \times 10^{-4}$. At longer times, arrest is seen as a weak plateau for the isotropic states and the nematic states $\Gamma = 0.16 - 0.25$, and as a clear plateau in the nematic state $\Gamma=0.29$ as well as in the liquid glass states $\Gamma = 0.31, 0.33$. At long times, particle motions again become diffusive for all states. Astonishingly, for large time scales $t>3 \times 10^2$ the liquid glass states $\Gamma=0.31, 0.33$ show identical diffusion independent of the density. The ellipsoids' centers of masses at different densities in the liquid glass for large timescales move similarly. This reminds of plots of the translation correlation functions in the liquid glass in the main article which show that the final decays of the translation correlation functions $F_s(q,t), S(q,t), S_{20}(q,t)$ at large timescales occur identically and independent of the liquid glass density (cf. Fig.14 and Fig.15 in the main article).


\end{document}